\documentclass[aps,prd,onecolumn,groupedaddress,nofootinbib,superscriptaddress]{revtex4}  
\usepackage{graphicx,mathtools,tabu}
\usepackage{epstopdf}
\usepackage{amsmath}
\usepackage{amsfonts}
\usepackage[colorlinks=true,citecolor=red,linkcolor=blue,breaklinks=true]{hyperref}
\usepackage{accents}
\usepackage[figure, table]{hypcap}
\usepackage{amssymb}
\usepackage{appendix}
\usepackage{comment}
\usepackage{bbold}
\usepackage{color}
\usepackage{slashed}
\usepackage{subfigure}
\usepackage{setspace}
\usepackage{footnote}
\usepackage{multirow}
\usepackage{braket}
\usepackage[normalem]{ulem}
\usepackage[utf8]{inputenc}
\usepackage{mathrsfs}
\usepackage{longtable}
\usepackage{enumitem}

\LTcapwidth=\textwidth

\usepackage{url}
\usepackage{wrapfig}
\usepackage{braket}

 \def\be   {\begin{equation}}  
 \def\ee   {\end{equation}}
 \def\ba   {\begin{array}}     
  \def\ea   {\end{array}}
 \def\bea  {\begin{eqnarray}}  
  \def\eea  {\end{eqnarray}}
 \def\bean {\begin{eqnarray*}}  
 \def\eean {\end{eqnarray*}}
 \def\ga   {\gamma}

  \def\al   {\alpha}
  \def\be {\beta}
    
      \def\r {\rho}

\def \R  {R_{\rm CCSN}}

\def\to {\rightarrow}

\newcommand{\eV}{{\rm\ eV}}

\newcommand{\MeV}{{\rm\ MeV}}

\newcommand{\Gpc}{{\rm\ Gpc}}

\begin{document}

 \hfill{NUHEP-TH/20-08, FERMILAB-PUB-20-353-T}
\title{Fundamental physics with the diffuse supernova background neutrinos}

\author{Andr\'{e} de Gouv\^{e}a}
\email{degouvea@northwestern.edu}
\affiliation{Northwestern University, Department of Physics \& Astronomy, 2145 Sheridan Road, Evanston, IL 60208, USA}
\author{Ivan Martinez-Soler}
\email{ivan.martinezsoler@northwestern.edu}
\affiliation{Northwestern University, Department of Physics \& Astronomy, 2145 Sheridan Road, Evanston, IL 60208, USA}
\affiliation{Theory Department, Fermi National Accelerator Laboratory, P.O. Box 500, Batavia, IL 60510, USA}
\affiliation{Colegio de F\'isica Fundamental e Interdisciplinaria de las Am\'ericas (COFI), 254 Norzagaray street, San Juan, Puerto Rico 00901.}
\author{Yuber F. Perez-Gonzalez}
\email{yfperezg@northwestern.edu}
\affiliation{Northwestern University, Department of Physics \& Astronomy, 2145 Sheridan Road, Evanston, IL 60208, USA}
\affiliation{Theory Department, Fermi National Accelerator Laboratory, P.O. Box 500, Batavia, IL 60510, USA}
\affiliation{Colegio de F\'isica Fundamental e Interdisciplinaria de las Am\'ericas (COFI), 254 Norzagaray street, San Juan, Puerto Rico 00901.}
\author{Manibrata Sen}
\email{manibrata@berkeley.edu}
\affiliation{Northwestern University, Department of Physics \& Astronomy, 2145 Sheridan Road, Evanston, IL 60208, USA}
\affiliation{Department of Physics, University of California Berkeley, Berkeley, California 94720, USA}

\begin{abstract}
The Universe is awash with tens-of-MeV neutrinos of all species coming from all past core-collapse supernovae. These have never been observed, but this state of affairs will change in the near future. In the less than ten years, the Super-Kamiokande experiment, loaded with gadolinium, is expected to collect dozens of events induced by the scattering of neutrinos from the diffuse supernova neutrino background (DSNB). Next-generation projects, including Hyper-Kamiokande and Theia, are expected to collect data samples with hundreds of DSNB events after a decade of running. Here, we study quantitatively how well the DSNB, including its energy spectrum, will be measured by different current or upcoming large neutrino detectors. We analyze the simulated data in order to estimate how well measurements of the DSNB can be used to inform research topics in cosmology -- including measurements of the Hubble parameter -- astrophysics -- including the star-formation-rate --  and particle physics -- including the neutrino lifetime and the possibility that neutrinos are pseudo-Dirac fermions.  
\end{abstract}

\maketitle

\section{Introduction}
A core-collapse supernova (CCSN) emits almost all its binding energy in the form of neutrinos. Observations of these neutrinos can provide crucial information on both the dynamics of CCSNe~\cite{Janka:2006fh,Horiuchi:2017qja} and the properties of neutrinos, including the neutrino lifetimes~\cite{Ando:2004qe}, magnetic moments~\cite{Ando:2003is,Balantekin:2007xq,deGouvea:2012hg}, and the number of neutrino species~\cite{Raffelt:2011nc}.
To date, we have only observed neutrinos from one CCSN: SN1987A~\cite{PhysRevLett.58.1490,PhysRevLett.58.1494}, in the Large Magellanic Cloud. In spite of the limited statistics, these provided priceless information and solidified the current understanding of the dynamics of CCSNe. 

Galactic CCSNe, however, are very rare; their rate is, on average, around one to three per century~\cite{Diehl:2006cf}. There is, however, an additional, strongly related but continuous source of astrophysical neutrinos: the diffuse supernova background (DSNB). The DSNB consists of neutrinos and antineutrinos emitted cumulatively from all past CCSNe in the observable Universe (see~\cite{Lunardini:2010ab,Beacom:2010kk} for a detailed review). The DSNB is expected to be, to a very good approximation, isotropic and time-independent. It can be determined from the rate of CCSNe in our Universe~\cite{2011ApJ...738..154H} and the flavor-dependent time-integrated neutrino spectra from CCSNe. We discuss the different ingredients that go into simulating the DSNB in Section~\ref{sec:model} and provide a brief summary in the next paragraphs.

The CCSN rate can be determined from observations of the cosmic star-formation rate (SFR), which can be  deduced from measurements of the distributions of stars as a function of their luminosities and masses~\cite{Hopkins:2006bw,Horiuchi:2008jz,Lunardini:2012ne}. With these observations, an analytical fit to the SFR can be obtained~\cite{Yuksel:2008cu,Horiuchi:2008jz}. This method depends on numerous astrophysical inputs, and is fraught with uncertainties~\cite{Nakazato:2015rya}. The neutrino flux from CCSNe is also not very well known, keeping in mind the very limited statistics from SN1987A~\cite{PhysRevLett.58.1490,PhysRevLett.58.1494}. One has to resort to large scale hydrodynamic simulations in order to estimate the energy and flavor spectra of neutrinos from CCSNe~\cite{Horiuchi:2017qja}. Different groups agree, with varying degrees of precision, that the neutrino spectra are quasi-thermal and can be described as pinched Fermi-Dirac spectra~\cite{Keil:2002in}. Finally, the neutrino fluxes are processed, deep inside the CCSN, by collective oscillations~\cite{Duan:2006an,Hannestad:2006nj,Fogli:2007bk,Dasgupta:2009mg,EstebanPretel:2007ec,Dasgupta:2008my,Dasgupta:2011jf,Chakraborty:2015tfa,Dasgupta:2016dbv,Izaguirre:2016gsx,Capozzi:2018clo,Bhattacharyya:2020dhu,Capozzi:2020kge,Johns:2020qsk} and, a little further out, by oscillations in the presence of ordinary matter with a decreasing number density. The latter are described well by the Mikheyev-Smirnov-Wolfenstein (MSW) resonant flavor conversion mechanism~\cite{PhysRevD.17.2369,Mikheev:1986gs}. Uncertainties in these different flavor conversion mechanisms also translate into uncertainties in the DSNB flux.

Previous studies have tried to address the uncertainties that plague the DSNB. A detailed study of the SFR~\cite{Horiuchi:2008jz}, incorporating multiple observations, confirmed the analytic fit and demonstrated that uncertainties in the SFR translate into an uncertainty of $40\%$ in the DSNB fluxes. On the neutrino-flavor-conversion side, a comprehensive study of the non-trivial effects of collective oscillations and MSW transitions on the DSNB neutrino flux have been performed in \cite{Chakraborty:2008zp,Chakraboty:2010sz,Moller:2018kpn}. Overall, non-trivial qualitative progress will only be made with a direct measurement of the DSNB.

Several publications have focussed on the prospects for the detection of the DSNB in current and future generation experiments including Super-Kamiokande (SK)~\cite{Zhang:2013tua}, Hyper-Kamiokande (HK)~\cite{Abe:2018uyc}, the Jiangmen Underground Neutrino Observatory (JUNO)~\cite{An:2015jdp}, and the Deep Underground Neutrino Experiment (DUNE)~\cite{Abi:2020evt}.
The sensitivity of the DSNB spectra to the local supernova-neutrino rate using a combination of CCSNe and failed supernovae have been studied in \cite{Moller:2018kpn}. 
 The strongest upper-limit on the DSNB flux, imposed by SK, hovers about a factor of a few above current predictions~\cite{Lunardini:2008xd,Zhang:2013tua}. Enrichment with gadolinium (Gd) will improve SK's sensitivity drastically, and is predicted to lead to the detection of the DSNB with three-sigma significance~\cite{Beacom:2003nk,Simpson:2018snj} in about five years. Future experiments like HK~\cite{Abe:2018uyc} and Theia~\cite{Askins:2019oqj}, a very large, water-based liquid-scintillator detector, have very good prospects for detecting the DSNB and, assuming a discovery in SK, measuring some of its properties.

Detecting the DSNB would open up new avenues in multi-messenger astronomy. In particular, it offers the unique opportunity to probe properties of the cosmos using neutrinos (as opposed to photons). Given current and near-future experimental capabilities, the sensitivity of neutrino observables is inferior to that of photons but the probes are qualitatively distinct and, in some circumstances, complementary. 
Measurements of the DSNB, for example, allow one to ``look back'' at the Universe with neutrinos and constrain the properties of its expansion. Instead, with precise knowledge of the expansion rate of the Universe, one can use the DSNB observation to inform the SFR. And finally, given external knowledge of the underlying cosmology and the SFR, one can probe new neutrino properties with the DSNB. In sum, the detection of the DSNB offers a unique way to pursue complementary avenues in particle astrophysics and cosmology using only neutrinos as probes (for example, see \cite{Barranco:2017lug,Moller:2018kpn,Jeong:2018yts,Yang:2019nnl,Dasgupta:2019cae,Bell:2020rkw}). 

In this work, we explore some these avenues. We simulate the DSNB event-rate in future neutrino detectors, particularly HK and Theia, and demonstrate how the DSNB can be utilized to inform the parameters of the Standard Model of Cosmology ($\Lambda$CDM)~\cite{Aghanim:2018eyx}, the SFR, or new neutrino properties. Details of our simulations are provided in Sec.~\ref{sec:detectors}. In Sec.~\ref{sec:cosmos}, we assume that there is no new physics beyond the Standard Model other than nonzero neutrino masses and study how well one can probe $\Lambda$CDM given our current knowledge of the SFR, and vice-versa. In Sec.~\ref{sec:new}, we ask the opposite question: given our current understanding of $\Lambda$CDM and the SFR, how well can the DSNB probe new neutrino properties? We consider two different scenarios: (i) neutrino decay, and (ii) pseudo-Dirac neutrinos. We find that the DSNB can provide very nontrivial information in both cases. We summarize our results in Sec.~\ref{sec:conc}.

\section{Modelling the DSNB flux}
\label{sec:model}
The DSNB flux is a function of the rate of CCSNe $(\R)$. This, in turn, depends on the history of star formation. An analytical fit of the redshift evolution of the co-moving SFR yields~\cite{Yuksel:2008cu,Horiuchi:2008jz}
\begin{equation}\label{eq:SFR}
    \Dot{\r}_*(z)=\Dot{\r}_0 \left[(1+z)^{-10\,\al}+
   \left( \frac{1+z}{B}\right)^{-10\,\be}+ \left(\frac{1+z}{C}\right)^{-10\,\ga}\right]^{-1/10}\,,
\end{equation}
where $\Dot{\r}_0$ is an overall normalization factor and $\al,\be,\ga$ are dimensionless exponents. 
The analytical fit to data is essentially a broken power-law, with $B$ and $C$ labeling the functional breaks at characteristic redshifts $z_1$ and $z_2$. They are defined as 
\begin{eqnarray}\label{eq:SFRPar}
B&=&(1+z_1)^{1-\al/\be}\,,\\
C&=&(1+z_1)^{(\be-\al)/\ga}(1+z_2)^{1-\be/\ga}\,.
\end{eqnarray}
The redshift breaks are defined to occur at $z_1=1$ and $z_2=4$. A fit to data from different astronomical surveys reveals $\Dot{\r}_0=0.0178^{+0.0035}_{-0.0036}\,{\rm M_\odot\,yr^{-1}\,Mpc^{-3}},\, \al=3.4\pm 0.2,\, \be= -0.3\pm 0.2$ and $\ga =-3.5\pm 1$.  
The variability in the comoving SFR considered in Ref.~\cite{Horiuchi:2008jz} takes into account the local rates estimated by different astronomical surveys. The negative-ten factors in Eq.~(\ref{eq:SFR}) are chosen to smoothen the fit. 

$\R$ also depends on the initial mass function (IMF) of stars $\psi(M)$, which gives the density of stars in a given mass range. The IMF is assumed to have a power-law behavior; according to  Salpeter,  $\psi\propto M^{-2.35}$~\cite{Salpeter:1955it}. With this information, 
\begin{equation}\label{eq:CCSR}
    \R(z)=\Dot{\r}_*(z)\frac{\int_8^{50} \psi(M)\,dM}{\int_{0.1}^{100} M \psi(M)\,dM}\,.
\end{equation}
The IMF in the numerator of Eq.~(\ref{eq:CCSR}) is integrated over the
mass range of $8\,M_\odot-50\,M_\odot$ to account for the stars that
can undergo a core-collapse, while the denominator accounts for all
stars up to $100\,M_\odot$. The fraction of stars undergoing a collapse to a SN corresponds to $0.007/M_\odot$. Stars with masses greater than
$50\,M_\odot$ lead to the formation of a black
hole~\cite{Moller:2018kpn} and are neglected in this analysis.

The diffuse differential neutrino flux, associated to an energy $E$ at the Earth, is 
\begin{equation}\label{eq:DSNB}
\Phi_{\nu}(E)=\int_0^{z_{\rm max}}\frac{dz}{H(z)}\R(z) F_\nu(E')\,,
\end{equation}
where 
\begin{align}
   H(z) = H_0\, \sqrt{\Omega_m (1+z)^3+\Omega_\Lambda(1+z)^{3(1+w)}+(1-\Omega_m-\Omega_\Lambda)(1+z)^2}
   \label{eq:Hubble}
\end{align}
is the Hubble function determined from the Friedmann equation. The Hubble parameter is taken to be, unless otherwise noted, $H_0=67.36\,{\rm km\,s}^{-1}\,{\rm Mpc}^{-1}$, in agreement with the best fit to the latest Planck data~\cite{Aghanim:2018eyx}. $z_{\rm max}\equiv5$ is the maximum redshift up to which there is a reasonable amount of star formation. Here, $\Omega_m$ and $\Omega_\Lambda$ refer to the matter and vacuum contributions to the energy density, in units of the total energy density, and $w$ is the equation-of-state parameter for the dark energy. $E'$ is the neutrino energy at the source, relative to which $E$ is redshifted,  $E=E'/(1+z)$.

$F_\nu(E)$ is the supernova neutrino spectrum. Most of the neutrino flux is emitted during the cooling phase of the supernova, which lasts for, very roughly, ten seconds post-bounce.  This phase is characterized by approximately equal luminosities for all neutrino flavors so each neutrino species has, to a good approximation, the same total energy $E_\nu^{\rm tot}/6$. 
Neutrino energy spectra from these CCSNe are approximately thermal and can be described by a non-degenerate time-integrated Fermi-Dirac distribution~\cite{Beacom:2010kk},
\begin{equation}\label{eq:Nuspec}
F_\nu(E)=\frac{E_\nu^{\rm tot}}{6}\,\frac{120}{7\pi^4}\,\frac{E_\nu^2}{T_\nu^4}\,\frac{1}{e^{E_\nu/T_\nu}+1}\,.
\end{equation}
$T_\nu$ is a species-dependent quantity as it is set by the neutrino interactions within the medium. Since temperatures inside the supernova allow, to a good-enough approximation, only the electron-flavored neutrinos to undergo charged-current interactions, the temperature hierarchy $T_{\nu_e}< T_{\bar{\nu}_e}<T_{\nu_x}$ is generically expected~\cite{Horiuchi:2017qja}. Here $\nu_x$ refers to either $\nu_\mu$ or $\nu_\tau$ or their antiparticles. The neutron richness of the supernova environment causes the $\nu_e$ to interact more than the $\bar{\nu}_e$, accounting for the lower value of $T_{\nu_e}$ relative to $T_{\bar{\nu}_e}$. These are the spectra at the so-called neutrinospheres (one for each species $\nu_e, \bar{\nu}_e,\nu_x$). Neutrino oscillations both inside the CCSN and between the CCSN and the Earth lead to ``mixed'' spectra at the time of detection. We will discuss our treatment of oscillation effects momentarily.

\begin{figure}[!t]
\includegraphics[width=.4\textwidth]{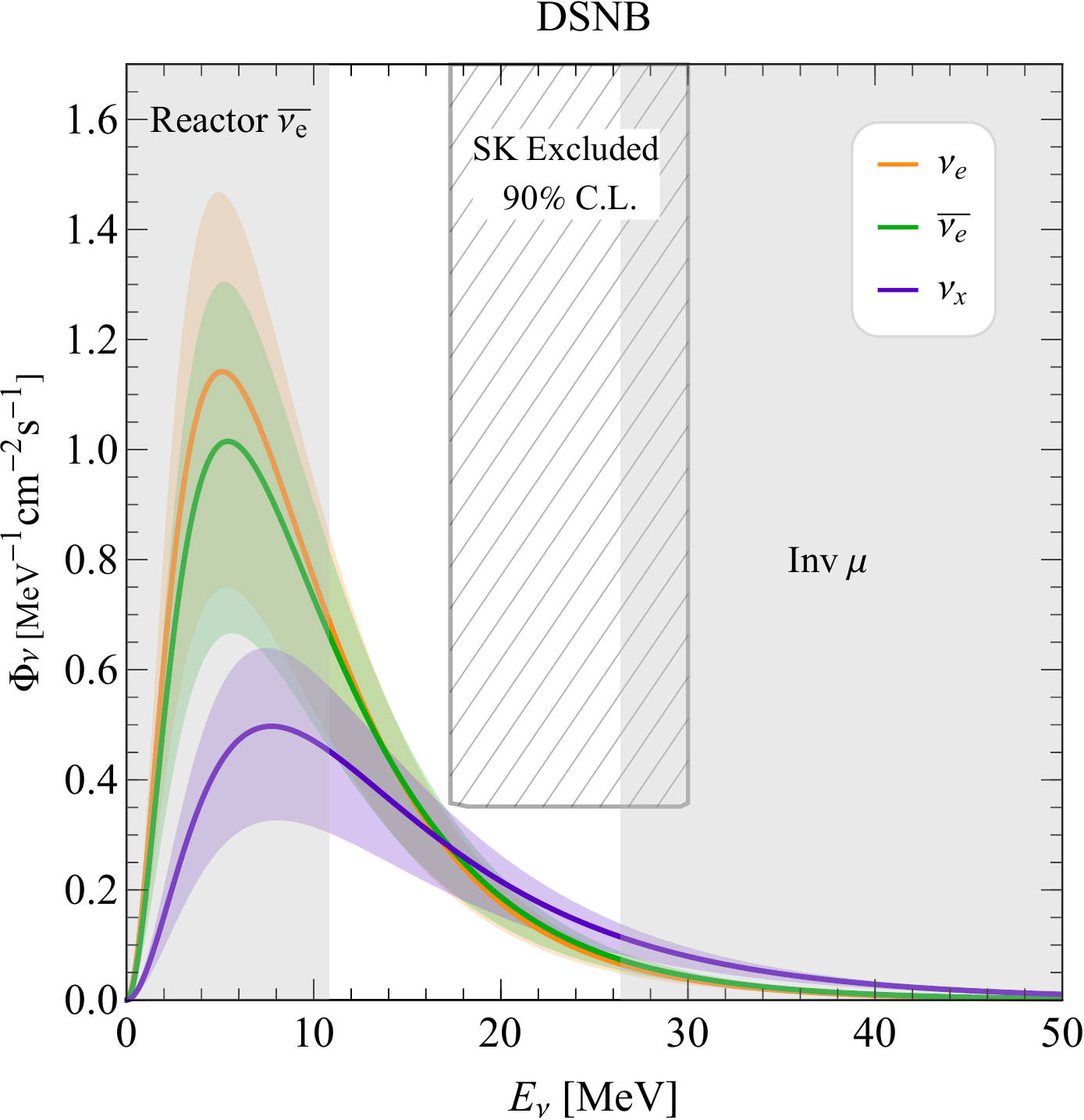}
\caption{The DSNB flux, excluding oscillation effects, for each neutrino species, as a function of the neutrino energy $E$ at the Earth. $T_{\nu_e}=6.6\,{\rm MeV},\, T_{\bar{\nu}_e}=7\,{\rm MeV}$ and $T_{\nu_x}=10\,{\rm MeV}$. The hatched region indicates current bounds from Super-Kamiokande \cite{Zhang:2013tua}.
The shading around the bands indicates the corresponding uncertainty arising from the SFR. The regions of energy-space where significant backgrounds are expected at the future facilities of interest are indicated with a gray background.}
\label{fig:DSNB}
\end{figure}
Fig.\,\ref{fig:DSNB} depicts the DSNB spectra as a function of $E$ for all three species of neutrinos, discounting oscillation effects. The shaded regions around the spectra indicate the corresponding uncertainty arising from the SFR. The hotter spectra extend out to higher energies while the cooler spectra have a more pronounced peak at lower energies. the hatched region indicates current bounds from Super-Kamiokande~\cite{Zhang:2013tua}. The regions of energy-space where significant backgrounds are expected at the future facilities of interest are indicated with a gray background. These will be discussed in detail in the next section.

Neutrino-oscillation effects inside the CCSN are significant and diverse; they depend, for example, on whether the neutrinos were produced earlier or later in the explosion. Neutrino interactions with the medium and neutrino interactions with other neutrinos lead to a rich oscillation pattern that is still the subject of intense theoretical and phenomenological investigation~\cite{Duan:2006an,Hannestad:2006nj,Fogli:2007bk,Dasgupta:2009mg,EstebanPretel:2007ec,Dasgupta:2008my,Dasgupta:2011jf,Chakraborty:2015tfa,Dasgupta:2016dbv,Izaguirre:2016gsx,Capozzi:2018clo,Bhattacharyya:2020dhu,Capozzi:2020kge,Johns:2020qsk}. Here, we ignore collective oscillations and assume that a strong, adiabatic MSW effect dominates neutrino flavor evolution for all energies of interest. We further assume that the neutrino mass ordering is, unless otherwise noted, known to be normal so all neutrinos that exit the neutrinosphere as $\nu_e$ will exit the CCSN as $\nu_3$ while all antineutrinos that exit the neutrinosphere as $\bar{\nu}_e$ will exit the CCSN as $\bar{\nu}_1$.\footnote{For the inverted neutrino mass ordering, still allowed by current data~\cite{Kelly:2020fkv}, all neutrinos that exited the neutrinosphere as $\nu_e$ would exit the CCSN as $\nu_2$ while all antineutrinos that exited the neutrinosphere as $\bar{\nu}_e$ would exit the CCSN as $\bar{\nu}_3$.} The other states exit the CCSN as incoherent mixtures of $\nu_1$, $\nu_2$, $\bar{\nu}_2$, and $\bar{\nu}_3$. In summary, we assume an incoherent mixture\footnote{In Sec.~\ref{sec:pseudo} we explore the hypothesis that neutrinos are pseudo-Dirac fermions and need to revisit this hypothesis more carefully.} of neutrino and antineutrino mass eigenstates arrives at the surface of the Earth. Their energy distributions are given by Eq.~(\ref{eq:Nuspec}) with $T_{\nu_3}=T_{\nu_e}$, $T_{\nu_2}=T_{\nu_1}=T_{\nu_x}$, $T_{\bar{\nu}_1}=T_{\bar{\nu}_e}$, and $T_{\bar{\nu}_2}=T_{\bar{\nu}_3}=T_{\nu_x}$. 

As we will discuss in the next section, we will concentrate on detectors that are sensitive to the  $\nu_e$ and $\bar{\nu}_e$ flavors. A mass eigenstate $\nu_i$ or $\bar{\nu}_i$ ($i=1,2,3$) arriving at the Earth's surface will be detected as a $\nu_e$ or a $\bar{\nu}_e$, respectively, with probability $P_{ei}$ and $\bar{P}_{ei}$. If the neutrinos do not travel a fair amount of the Earth in order to get to the detector, $P_{ei}=\bar{P}_{ei}=|U_{ei}|^2$ ($i=1,2,3$), where $U_{ei}$ are elements of the leptonic mixing matrix. Instead, if the neutrinos have to travel a fair amount of Earth in order to reach the detector, Earth matter effects need to be taken into account. Assuming the DSNB flux is isotropic, these effects can be included via the average probabilities (as a function of the neutrino energy) $\langle P_{ei}\rangle$ and $\langle\bar{P}_{ei}\rangle$. We checked that the impact of Earth matter-effects is safely negligible for our purposes and have not included them here. 

It turns out that neutrino oscillation effects have a small impact in the results we discuss here. In the energy range of interest (see Fig.~\ref{fig:DSNB}), the different $F_\nu(E)$ for the different neutrino species are rather similar given the large SFR uncertainties. Had they been exactly the same, oscillation effects would have been completely absent so the impact of the oscillations, very generically, are proportional to the differences among the initial spectra. Since that these are small given the large uncertainties, oscillation effects are also small. 

\section{Detecting the DSNB}
\label{sec:detectors}
The DSNB flux is very small, concentrated around neutrino energies of order tens of MeV.  Large neutrino detectors sensitive to these neutrino energies have to contend with several different sources of background~\cite{Lunardini:2010ab} that render the unambiguous detection of the DSNB a difficult task~\cite{Zhang:2013tua}. It is, however, widely expected that the next generation of experiments will be both large and well-equipped enough to identify signal from background to detect the DSNB in the next five to ten years~\cite{Beacom:2003nk,Simpson:2018snj}. Indeed, several upcoming or planed experiments should be able to not only detect but also measure some of the properties of the DSNB. These include, in the near future, Super-Kamiokande~\cite{Simpson:2018snj} and JUNO~\cite{An:2015jdp} and, further down the road, the Hyper-Kamiokande~\cite{Abe:2018uyc} and Theia~\cite{Askins:2019oqj} detectors. 

In order to estimate the sensitivity to the DSNB, we compute the expected event rate for each experiment as a function of the reconstructed energy. Those events are distributed in an array of bins of reconstructed energy, the size of the bin depending on the energy resolution of each proposed experiment. Inside a given bin \textit{i}, the expected total number of events is
\begin{equation}
    N_{i} = N_{\rm tar} T \int dE^{r}dE^{t}\Phi_{\alpha}\sigma_{\alpha}\epsilon(E^{t},E^{r}) + \text{Bkg}_{i}\,,
\end{equation}
where $\sigma_{\alpha}$ is the cross-section of the main detection channel, $\alpha=e,\mu,\tau$ is a flavor index, $\Phi_{\alpha}$ is the flux of each neutrino species at the detector, and $\epsilon(E^{t},E^{r})$ correlates the true neutrino energy $E^t$ with the experimentally reconstructed energy $E^r$. The integrals are over the values of $E^r$ that define the bin and over all $E^t$. The normalization constants $N_{\rm tar}$ and $T$ correspond to the number of targets in the detector and the running time, respectively. Here, we assume 10~years of data taking unless otherwise noted. The correlation matrix $\epsilon$ for each experiment was obtained by Montecarlo integration. $\text{Bkg}_{i}$ is the number of expected background events in the $i$th energy bin after all background-reduction information has been applied. These also depend, of course, on the type of experiment.

The dominant background to detect the DSNB depends on the energy. For energies above $\sim 40$~MeV, atmospheric neutrinos make up the majority of the neutrino flux~\cite{Battistoni:2005yu}. Atmospheric muons also contribute to the background via muon-spallation \cite{Li:2014sea,Abe:2018uyc}. At lower energies, neutrinos from the Sun or nearby nuclear reactors become a significant component of the neutrino flux. For energies below $10$~MeV, these are expected to overwhelm the DSNB. 

We will concentrate on the detection of $\bar{\nu}_e$ via inverse beta-decay (IBD) in water and liquid scintillator experiments and on the detection of $\nu_e$ in liquid Argon experiments, discussed later. For neutrino energies typical of the DSNB, the main detection channel for electron antineutrinos is the IBD interaction with free protons, $\bar{\nu}_{e} + p \rightarrow n + e^{+}$~\cite{Strumia:2003zx}; unless otherwise noted, we ignore all $\bar{\nu}_e$ subdominant detection channels in the experiments we will be considering here.  The identification of the positrons created in the IBD interactions of the DSNB is challenging due to the large number of background sources of electrons or positrons in the same energy range, around tens of MeV. For energies below 10~MeV, the dominant irreducible background consists of $\bar{\nu}_e$ from nuclear reactors. We assume those overwhelm the DSNB for all experimental setups of interest. For higher energies, the reactor flux is significantly smaller than the expected DSNB flux and the main background sources are muon-spallation products, atmospheric $\bar{\nu}_e$ -- this background is irreducible -- the decay of invisible muons produced in the charged current interactions of atmospheric muon neutrinos and antineutrinos, and neutral-current scattering of atmospheric neutrinos of all species. Some of these backgrounds are, however, reducible if one is able to identify the neutrons produced in the IBD process: neutron tagging. These neutrons are associated with the IBD positron both in time and space. Efficient neutron tagging is fundamental for the detection of $\bar{\nu}_e$ from the DSNB.
 
In what follows, we describe the main techniques proposed for the measurement of the DSNB, focusing in the different detector technologies that will be used by future experiments and projects. 

\emph{Water doped with gadolinium -- SK and HK} --- Starting this calendar year, Super-Kamiokande (SK), a 50~kton water Cherenkov detector running in Japan, will be doped with gadolinium. The main purpose of the gadolinium doping is to increase SK's sensitivity to the DSNB~\cite{Beacom:2003nk}. Other immediate applications include precision measurements of the flux of antineutrinos produced in nuclear reactors around SK~\cite{Beacom:2003nk,deGouvea:2020vww}. In the future, Hyper-Kamiokande (HK), another water Cherenkov detector proposed in Japan, with a fiducial volume of 187~kton~\cite{Abe:2018uyc}, is also expected to be doped with gadolinium. 

In water, neutrons are absorbed by hydrogen, emitting a photon with $2.2$~MeV. Using neutron tagging with hydrogen, SK has searched for the DSNB flux~\cite{Zhang:2013tua}. Remaining backgrounds were such that SK was restricted to look for the DSNB in the energy window depicted in Fig.~\ref{fig:DSNB}. Unlike hydrogen, gadolinium has a large cross-section for capturing neutrons and these are associated with the emission of  $8$~MeV of photons. Both of these facts increase the neutron-tagging capabilities of large water Cherenkov detectors. The impact of the neutron tagging with Gd is expected to reduce the background signal below that of the expected DSNB flux for energies between $10$~MeV and $~40$~MeV. The reduction is particularly significant for the muon-spallation background, and only spallation nuclei created together with a neutron, like $^{9}$Li, may still be misidentified as an IBD signal. It is expected that with the Gd-doping, SK will be sensitive to the DSNB for antineutrino energies immediately above the reactor background ($E\gtrsim 10$~MeV).  For antineutrino energies above, roughly, 40~MeV, the invisible antimuon-decay background overwhelms the DSNB signal.

\begin{figure}[!t]
\includegraphics[width=0.75\textwidth]{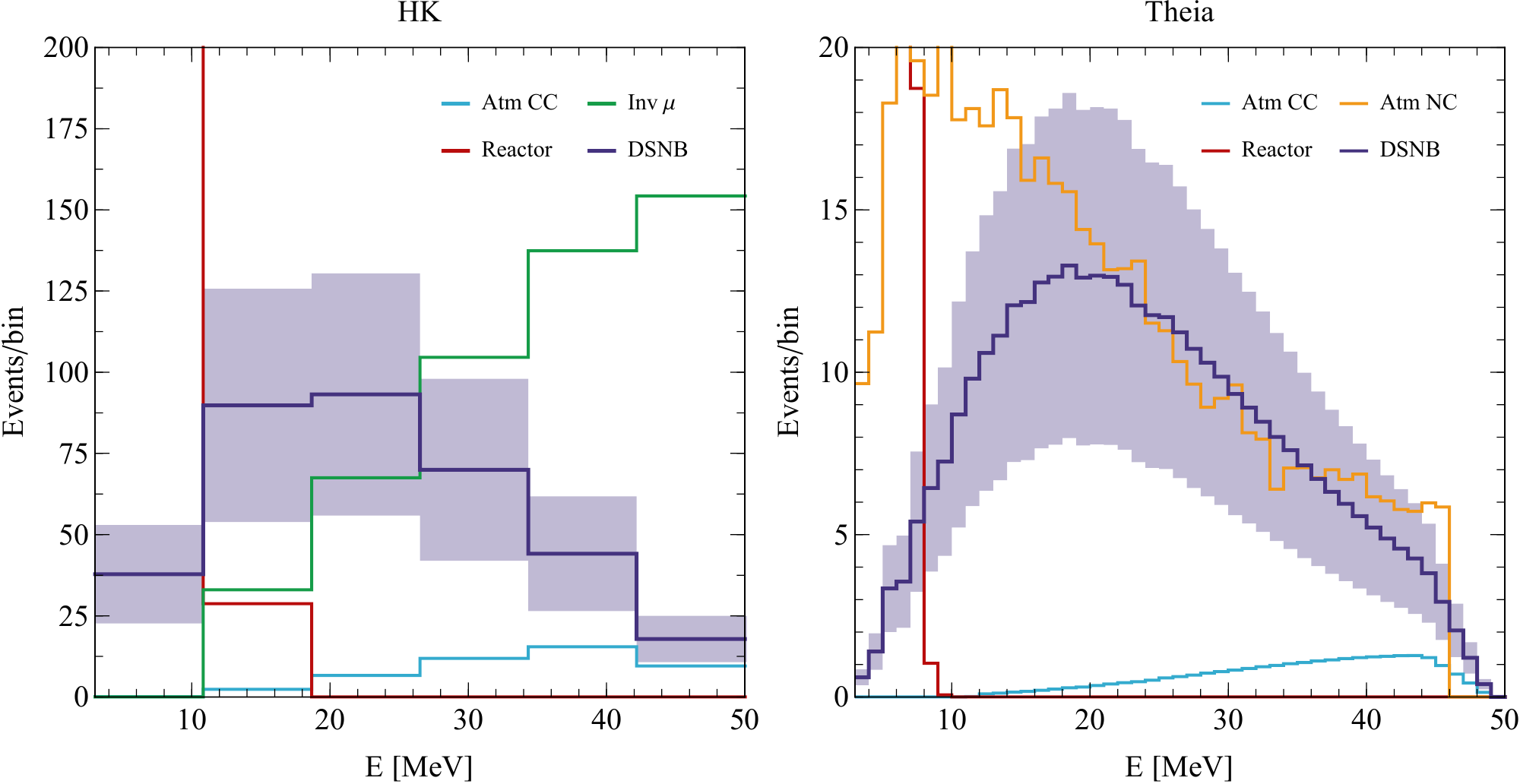}
\caption{Number of events expected for HK (left) and Theia (right) in 10 years, from different sources. The DSNB is in purple, this shading indicating the corresponding modeling uncertainty, while different background sources correspond to differently colored histograms. For HK, background sources are reactor antineutrinos (red), atmospheric $\bar{\nu}_e$ (cyan) and the decay of invisible atmospheric-muons (green). For Theia, the background sources are atmospheric $\bar{\nu}_e$ (cyan), neutral-current atmospheric neutrino scattering on carbon (orange) and reactor antineutrinos (red). Some of these backgrounds are reducible. See text for details.}
\label{fig:EventHK}
\end{figure}

In order to estimate the sensitivity of SK and HK to the DSNB flux, we compute the number of events for neutrinos in the energy range from $10$~MeV to $50$~MeV. We distributed those events in energy bins of $8$~MeV; the size of the bins is two times the uncertainty in the energy reconstruction at the maximum energy considered. We assume the energy resolution to be $\sigma_{E} = 0.6\sqrt{E/{\rm MeV}}$, corresponding to an uncertainty of $20\%$ at 10~MeV~\cite{Abe:2018uyc} for both SK and HK. Fig~\ref{fig:EventHK} (left) depicts the number of events expected for HK (purple) together with the contribution of the different backgrounds. The shading indicates the uncertainly associated with the SFR.

\emph{Liquid Scintillator (LS) -- JUNO} --- Organic liquid scintillators are another popular medium for the detection of electron antineutrinos with tens of MeV of energy via IBD. JUNO is a $20$~kton LS detector under construction in China and expected to be the largest LS detector for the foreseeable future. Its main mission is to precisely measure the flux of electron antineutrinos from nearby nuclear reactors. The nuclear power plants at Taishan and Yanjiang, located $53$~km from JUNO, limit the detection of the DSNB flux to energies above $10$~MeV. At higher energies, the main backgrounds come from atmospheric muons and neutrinos. Backgrounds from muon-spallation products in LS detectors dominate below $10$~MeV and are less significant in the energy range of interest. In addition to the charged-current (CC) interactions of atmospheric $\bar{\nu}_{e}$ (irreducible background), the neutral-current (NC) interactions of  high-energy atmospheric neutrinos with the carbon nuclei can lead to the emission of a neutron and to an excited nucleus that will emit $\gamma$-rays as it de-excites~\cite{Collaboration:2011jza}. That signal will mimic the IBD events of interest. The NC background can be reduced using timing information of the de-excitation of the carbon nucleus and the shape of the signal~\cite{Mollenberg:2014pwa}. LS is more efficient in the identification of muons than water Cherenkov experiments, thanks to the detection of the Michel electron associated to the muon decay at rest and to more information on the shape of the event. This allows one to effectively eliminate all backgrounds associated to muons interacting or decaying inside the detector. Muons that interact outside the detector, however, can produce fast neutrons that can penetrate the detector volume and interact with protons. The recoil of the proton and the absorption of the neutron will lead to a signal that can mimic the IBD. A shape-analysis and reduction in the fiducial volume -- the short penetration length of the neutrons implies that those events are concentrated towards the boundary of the detector -- will reduce this background.   

When simulating the JUNO detector, we consider a fiducial volume of 17~kton. For energy resolution, we consider an uncertainty $\sigma_{E}=0.03\sqrt{E/{\rm MeV}}$. We consider the DSNB flux from $10$~MeV to $50$~MeV, distributing the events in bins of $1$~MeV.

\emph{Water-based Liquid Scintillator (WbLS) -- Theia} --- Some liquid scintillating materials can be dissolved in water. The detection material, therefore, consist of a solution of $1\%-10\%$ liquid scintillator in water. Theia is a proposal for one of the four modules in the DUNE far detector complex and plans to use WbLS as its detector material. There are two proposals for the size-configuration, 25~kton and 100~kton. The results presented here are based on the larger realization. The advantage of using this kind of target material is the possibility to combine the reconstruction techniques developed for water Cherenkov and liquid scintillator detectors. In a nutshell, water Cherenkov detectors are good at reconstructing event topologies whereas the liquid scintillator provides precise information on the energy of the event. In addition to the irreducible backgrounds -- neutrinos from reactors at low energies (below $\sim 10$~MeV) and atmospheric antineutrinos at high energies (above $\sim 40$~MeV) --  similar backgrounds to those at LS detectors will affect WbLS detectors. In the energy region of interest, the dominant backgrounds include backgrounds of atmospheric origin including muon-spallation $^9$Li, the NC interactions of atmospheric neutrinos, or the fast neutrons created outside the detector that penetrate and interact elastically with free protons. Studying the topology and the time distribution of each event, together with the ratio between the Cherenkov and scintillation light, WbLS detector appear to be  capable of eliminating most of these sources of background, reducing remaining sources by more than one order of magnitude. The sources that cannot be eliminated are the atmospheric CC $\bar{\nu}_e$ and NC neutrino interactions~\cite{Gann:2015fba,Askins:2019oqj}. 

When simulating events at Theia, we consider a fiducial volume of $80$~kton and $6.2\times 10^{33}$ targets. We compute the number of events in the energy range from $10$~MeV to $50$~MeV using bins of $1$~MeV. For energy resolution, we assume $\sigma_{E} = 0.03\sqrt{E/\text{MeV}}$. The expected event distribution (purple) as well as the main background expectations are depicted in Fig.~\ref{fig:EventHK} (right). The energy resolution for Theia will depend on the fraction of scintillator dissolved in the water. The value we use here corresponds to the most optimistic expected energy resolution. 

\emph{Liquid Argon (LAr) -- DUNE} --- Liquid-argon detectors are currently the subject of intense experimentation. The DUNE far detector~\cite{Abi:2020evt}, which will be ready before the end of decade, is expected to be the largest such detector for the foreseeable future with 40~ktons of LAr. LAr is mainly sensitive to the $\nu_{e}$ component of the DSNB flux. These are measured via neutrino absorption in $^{40}$Ar, yielding an electron and an excited state of potassium, $\nu_e + ^{40}{\rm Ar} \rightarrow ^{40}{\rm K}^{*} + e^{-}$. In the de-excitation process, the potassium nucleus emits a cascade of photons. Both the electron and the photons are measured as MeV blips in the LAr. The dominant backgrounds here -- mostly irreducible -- include the flux of $\nu_{e}$ from the atmosphere, for energies above $40$~MeV, and the solar neutrinos, for energies below $16$~MeV.

When simulating events in the DUNE far detector, we consider a fiducial volume of $40$~ktons. To simulate the interaction of the DSNB flux with LAr, we used MARLEY~\cite{Gardiner2018}, a Montecarlo event generator that provides the final states and their energies after the neutrino interaction for a given neutrino energy. We assume that electrons and photons are observed if their energies are above $2$~MeV. Regarding the energy resolution, we assume the same precision demonstrated in previous LAr experiments~\cite{Amoruso:2003sw}, $\sigma_{E}=0.11\sqrt{E/\text{MeV}} + 0.2 (E/\text{MeV})$. The expected events are distributed in energy bins of 5~MeV.

For the analysis of all simulated data, we include systematic uncertainties associated to the DSNB flux in the form of a 40\% uncertainty in the overall normalization and include uncertainties associated to the current knowledge of the leptonic mixing matrix from the results of the global fit in Ref.~\cite{Esteban:2018azc}. The latter have a negligible impact in our results, presented in the next two sections.

\section{Cosmology and astrophysics potential of the DSNB}
\label{sec:cosmos}
Eq.\,(\ref{eq:DSNB}) reveals that the DSNB fluxes (different flavors and energies) broadly depend on three input quantities: 
\begin{enumerate}
\item the rate of core-collapse supernovae $\R$, which in turn depends on the SFR. This depends on various astrophysical inputs, as discussed earlier;
\item the cosmological history of the Universe, through the parameters  $\Omega_m$, $\Omega_{\Lambda}$, and $H_0$, assuming the $\Lambda$CDM model;
\item the spectra of the emitted neutrinos $F_\nu(E)$.
\end{enumerate}
These three ingredients are entwined in the observables related to the DSNB we can hope to measure in the future, i.e., energy spectra inside a limited energy window for different neutrino species. Assuming external knowledge on subsets of these parameters, however, measurements of the DSNB can be used to inform the complementary subset in a nontrivial way. 

In this section, we look at the first two directions. We assume that there is no exotic physics in the neutrino sector, i.e., the only new physics we assume is that neutrinos have nonzero masses and oscillate, so the input that goes into $F_\nu(E)$ is as described in Eq.\,(\ref{eq:Nuspec}) and Sec.~\ref{sec:model}. In Sec.~\ref{sec:LCDM}, we assume that the SFR is indeed given by Eq.\,(\ref{eq:SFR}) and constrained as discussed in Sec.~\ref{sec:model}, and try to constrain some of the different parameters of $\Lambda$CDM cosmology using the simulated events. In Sec.~\ref{sec:SFR}, we ask the reverse question: if cosmology is indeed governed by $\Lambda$CDM with the different parameters well-determined by Planck \cite{Aghanim:2018eyx} and other cosmic surveys, how well can we constrain the SFR parameters? 
\subsection{Probing $\Lambda$CDM Cosmology with the DSNB}
\label{sec:LCDM}
The DSNB, composed of neutrinos coming from all possible CCSNe at redshifts $z\lesssim 5$, is sensitive to the low redshift expansion of the Universe. Therefore, measurements of the DSNB serve as unique tools to constrain the underlying six-parameter $\Lambda$CDM cosmology using neutrinos. While, for example, measurements of different properties of the cosmic microwave background (CMB) already provide excellent measurements of these parameters~\cite{Aghanim:2018eyx}, these measurements are essentially properties of the high-redshift $(z\simeq 1100)$ Universe. The DSNB offers a very distinct probe of the same parameters at low redshifts. In particular, with the tension between the extracted value of the Hubble parameter $H_0$ inferred from CMB data~\cite{Aghanim:2018eyx} and those from low redshift measurements of Type Ia supernovae~\cite{Riess:2016jrr,Bernal:2016gxb,Birrer:2018vtm,Riess:2019cxk}, an independent, low-redshift measurement of $H_0$ with neutrinos from CCSNe would play a  role in resolving the tension. While the Planck measurements are expected to be much more precise owing to the high statistics of CMB photons, future measurements of the DSNB offer an intriguing, albeit, a lot less precise, way of looking at the early Universe using neutrinos.

\begin{figure}[!t]
\includegraphics[width=.75\textwidth]{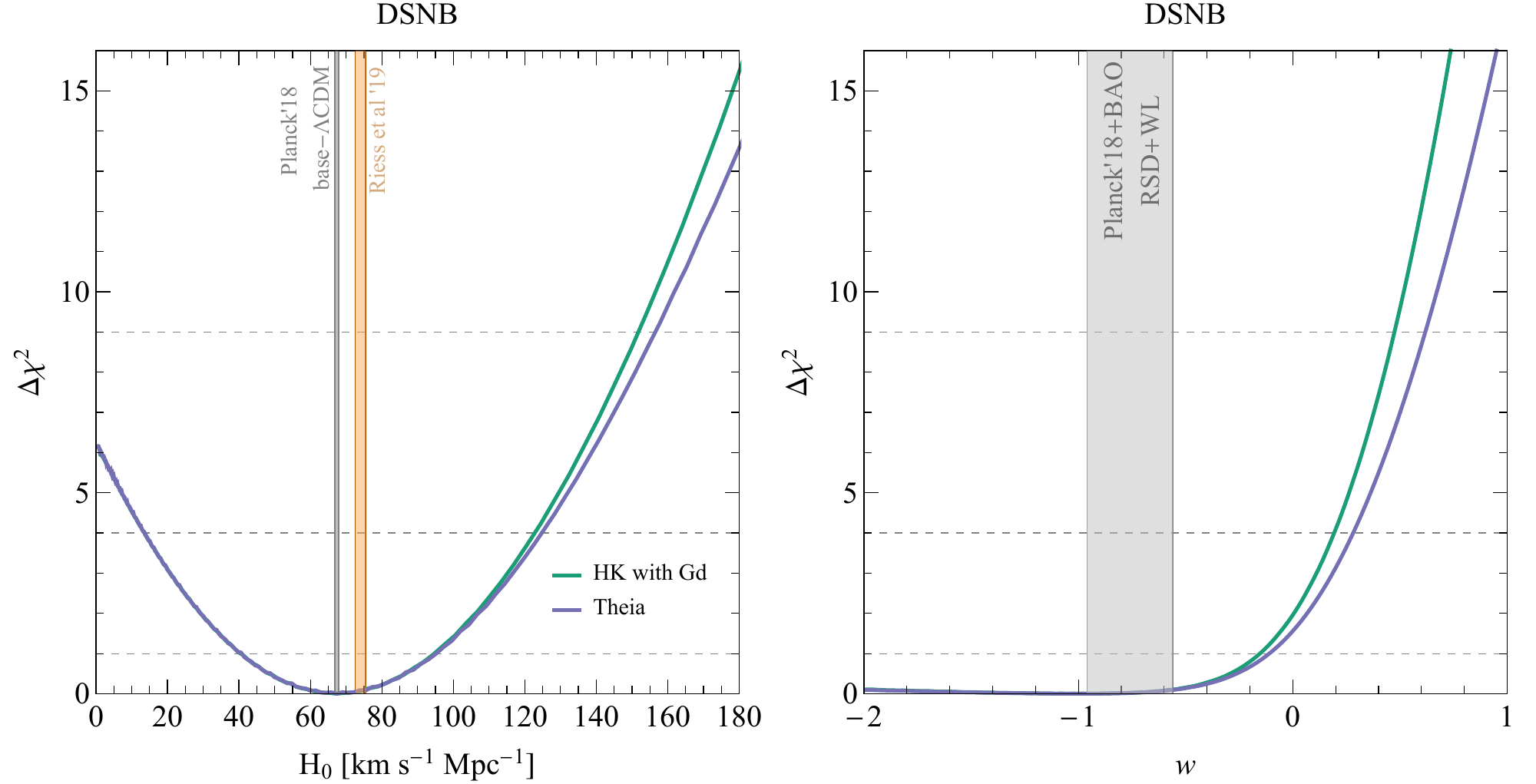}
\caption{$\Delta\chi^2$ (relative to the minimum value) as a function of the Hubble parameter $H_0$ (left) and the equation of state parameter for the dark energy $w$ (right), for ten years of simulated data at Theia (blue) and HK enriched with Gd (green). The vertical bands in the left-panel indicate the current bound from CMB observations by Planck (gray)~\cite{Aghanim:2018eyx} and observations at low redshifts (orange)~\cite{Riess:2016jrr,Bernal:2016gxb,Birrer:2018vtm,Riess:2019cxk}. The vertical band in the right-panel indicates the current bound from Ref.~\cite{Aghanim:2018eyx}.}
\label{fig:cosmo1d1}
\end{figure}

The left panel of Fig.\,\ref{fig:cosmo1d1} depicts the sensitivity of HK and Theia to the Hubble parameter, $H_0$, measured using the DSNB. Since $H_0$ controls only the absolute normalization to the DSNB flux, it is more sensitive to the number of events than their energy distribution. In the fit, we assume the parameters associated with the SFR and neutrino mixing are known perfectly and agree with their current best-fit values. As discussed earlier, the 40\% systematic uncertainty assumed for the overall neutrino flux partially takes care of uncertainties associated to the SFR and neutrino mixing. 
Both HK and Theia can rule out, at more than the two sigma level, $H_0<10\, {\rm km\,s}^{-1}\,{\rm Mpc}^{-1}$ and can hence confirm that the universe is expanding. On the other hand, very large values of $H_0$ can be ruled out at more than the three-sigma level. As advertised, DSNB measurements are not competitive with those using photons at either high~\cite{Aghanim:2018eyx} or low redshifts~\cite{Riess:2016jrr,Bernal:2016gxb,Birrer:2018vtm,Riess:2019cxk}. Nonetheless, it is remarkable that one can obtain a nontrivial, independent measurement of $H_0$ using the DSNB flux.\footnote{Strictly speaking, we are measuring the impact of the expansion rate of the universe on the propagation of neutrinos between the source and the detector. In our analysis, we assume that these effects are divorced from the impact of the expansion rate when it comes to determining the SFR, for example.} The figure reveals that after ten years, both HK and Theia would be able to measure $H_0$ at the 40\% level. Given the 40\% systematic uncertainty on the expected flux assumed here, this means the ten-year result is already systematics dominated. A significantly more precise measurement would require qualitatively better understanding of the theoretical expectations for the DSNB flux.

The DSNB signal is not very sensitive to the energy density parameters $\Omega_{m,\Lambda}$; it can, however, be used to constrain the Dark Energy equation-of-state parameter $w$. 
The right panel Fig.\,\ref{fig:cosmo1d1} depicts the sensitivity of HK and Theia to $w$. Assuming the true value of $w=-1$ (cosmological constant), both HK and Theia can exclude $w\gtrsim 0.5$ at around the three-sigma level. Due to higher statistics, HK does a marginally better job than Theia.

\subsection{Probing the star-formation rate with the DSNB}
\label{sec:SFR}
One of the major inputs that go into calculating the DSNB spectra is the rate of CCSNe in the observable Universe. $R_{\rm CCSN}$ is governed by the star-formation ratio, given by Eq.\,(\ref{eq:SFR}). Data from different astronomical surveys provide key insights into the cosmic SFR up to redshifts $z\simeq 6$~\cite{Hopkins:2006bw,Horiuchi:2008jz,Lunardini:2012ne}. The uncertainties arising from these surveys translate into uncertainties in the fit parameters $(\Dot{\r}_0,\al,\be,\ga)$, as outlined in Sec.\,\ref{sec:model} (also see Table I of Ref.~\cite{Horiuchi:2008jz}). 

\begin{figure}[!t]
\includegraphics[width=.75\textwidth]{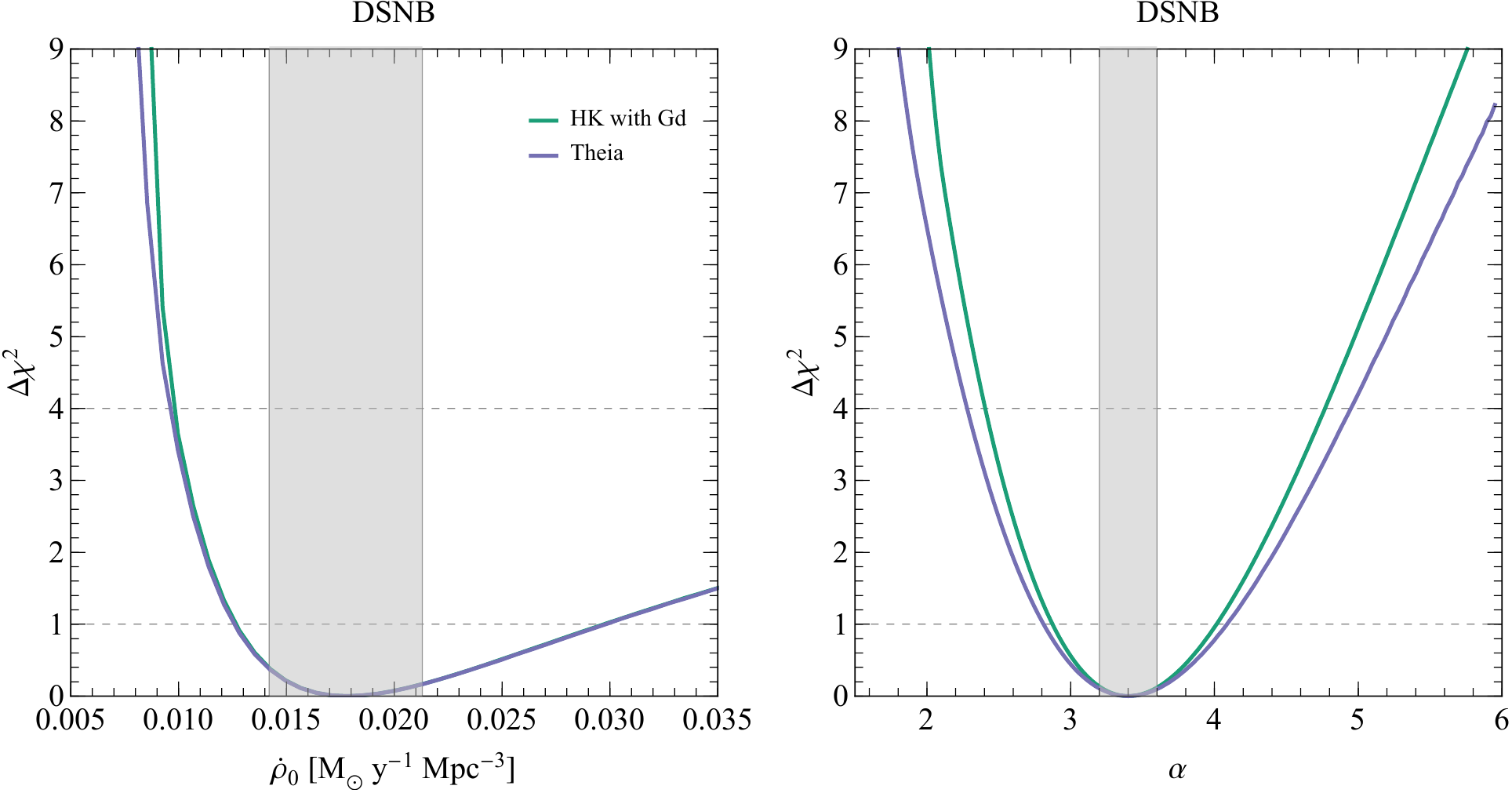}
\caption{$\Delta\chi^2$ (relative to the minimum value) as a function of $\Dot{\r}_0$ (left) and $\alpha$ (right) for HK with Gd (green), and Theia (blue). The vertical shaded regions indicate the current allowed range for the relevant parameters, obtained from fits to astronomical observations \cite{Hopkins:2006bw,Horiuchi:2008jz}. See text for details.}
\label{fig:rccsn1d}
\end{figure}

It is interesting to consider the prospects of constraining the SFR parameters through the detection of the DSNB neutrinos in HK~\cite{Mathews:2014qba} and Theia~\cite{Askins:2019oqj}. A recent discussion of the capabilities of HK can be found in Ref.~\cite{Riya:2020wpw}. We simulate events in these detectors, assuming the SFR is described by the current best-fit values of $(\Dot{\r}_0,\al,\be,\ga)$, and allow these parameters to vary one at a time when analyzing the simulated data. In the fit, we assume the parameters associated with the expansion rate of the universe and neutrino mixing are known perfectly and agree with their current best fit value. The same is true of the SFR parameters not under investigation. As discussed earlier, the 40\% systematic uncertainty assumed for the overall neutrino flux partially includes the uncertainties associated to the SFR, cosmological parameters and neutrino mixing. We find that the DSNB measurements are most sensitive to $\Dot{\r}_0$ and $\al$, and these parameters can be non-trivially constrained. 
Fig.\,\ref{fig:rccsn1d} depicts the $\Delta \chi^2$ as a function of $\Dot{\r}_0$ (left panel), and $\al$ (right panel). $\Dot{\r}_0$ functions as an overall normalization, and both HK and Theia have similar sensitivity to it, as we observed in the $H_0$ discussion in the previous subsection. On the other hand, when it comes to constraining $\al$, HK has slightly better sensitivity than Theia. The bounds arising from fits to the astronomical survey data, presented in Sec.~\ref{sec:model}, are depicted in gray.
At the two-sigma level, the results obtained from the measurement of the DSNB are almost competitive with those obtained from astronomical surveys. 

\begin{figure}[!t]
\includegraphics[width=.75\textwidth]{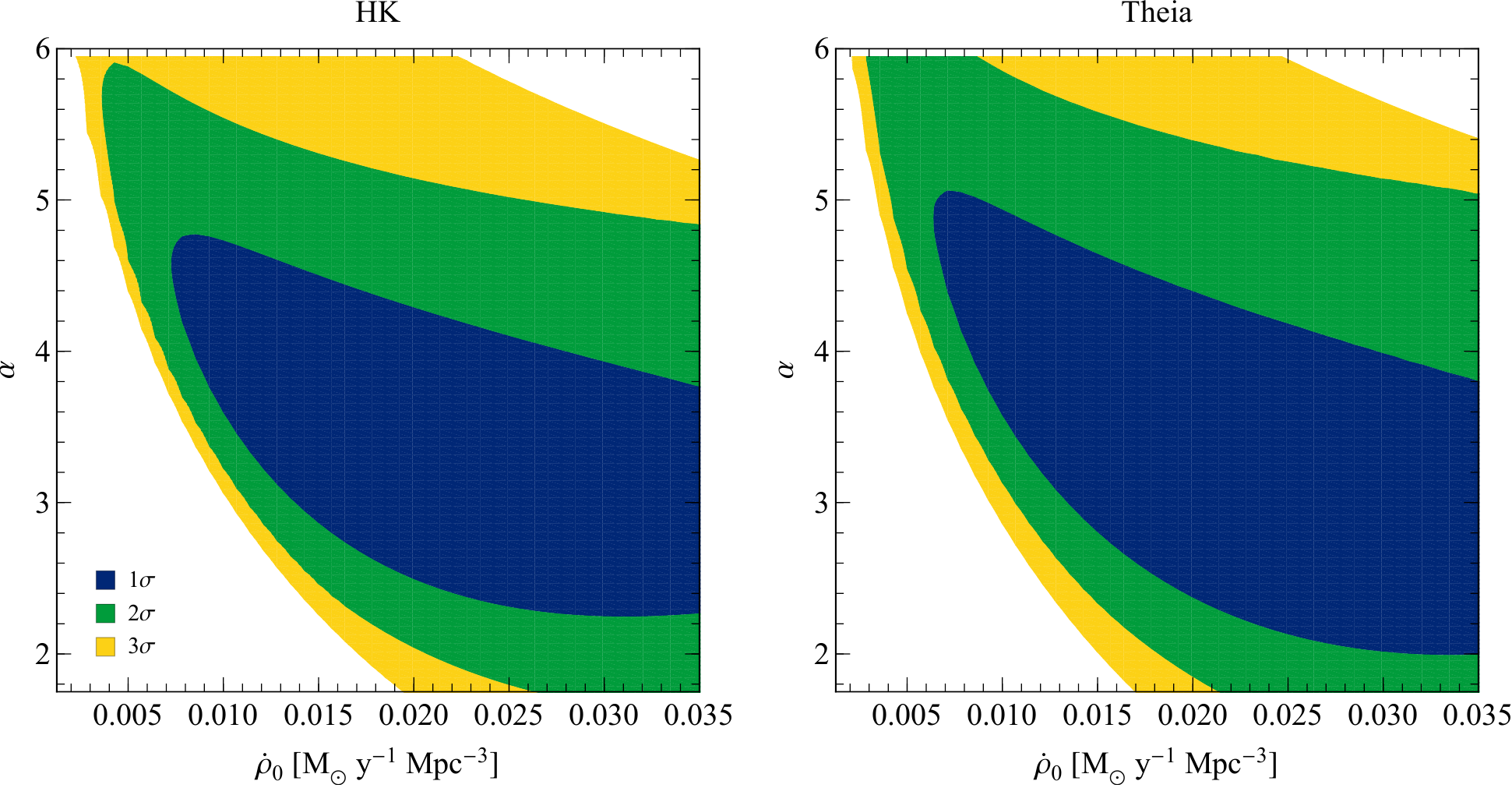}
\caption{Allowed regions of the $\Dot{\r}_0$ vs $\alpha$ parameter space, from ten years of HK with Gd (left) and Theia (right), at the $1,2,3-\sigma$ confidence levels. See text for details.}
\label{fig:rccsn2d}
\end{figure}

We also studied the sensitivity of the DSNB signal to $\Dot{\r}_0$ and $\al$ simultaneously. Fig.\,\ref{fig:rccsn2d} depicts the allowed contours in the $\Dot{\r}_0-\al$ plane for HK (left panel) and Theia (right panel). Both HK and Theia exhibit a clear correlation between these two SFR parameters. With access to more data, one can expect the contours to shrink, especially in the $\alpha$ direction. As before, in the fit, all other parameters are held fixed. 

\section{Constraining exotic neutrino physics with the DSNB}
\label{sec:new}
In the previous section, we demonstrated how the DSNB can be used to put constraints on cosmological as well as astrophysical parameters, assuming one has a clear understanding of the underlying neutrino physics. However, one might turn the argument around and use the DSNB as a probe of exotic physics in the neutrino sector. This, of course, requires knowledge of the underlying cosmological model and the star-formation rate. In this section, we assume that the underlying cosmology is governed by $\Lambda$CDM and that the vacuum and matter energy density fractions are well-constrained by Planck~\cite{Aghanim:2018eyx} and other cosmic surveys. Furthermore, we assume the SFR is governed by Eq.\,(\ref{eq:SFR}) and that the phenomenological parameters are within the range quoted in Sec.~\ref{sec:model}.

One can use the naturally long baseline offered by the DSNB to constrain ``slow'' neutrino dynamical processes. These include testing the hypothesis -- from Standard-Model-only interactions and massive neutrinos --  that the neutrino lifetime is much longer than the age of the Universe, and looking for new neutrino oscillation-lengths associated to tiny new mass-squared differences. The latter searches allow one to constrain the hypothesis that massive neutrinos are pseudo-Dirac fermions. 

\subsection{Probing neutrino decays with the DSNB}
Massive neutrinos can decay into lighter ones. Within the SM, these decays take place via weak electromagnetic-dipole-diagrams and weak box-diagrams, and the expected lifetimes are much longer than the age of the Universe. However, if the neutrino has interactions beyond those in the SM then it can decay faster. Such non-radiative decays of the neutrino have been studied extensively, in the context of terrestrial and astrophysical neutrinos~\cite{Berezhiani1992,Fogli:1999qt,Choubey:2000an,Lindner:2001fx,Beacom:2002cb,Joshipura:2002fb,Bandyopadhyay:2002qg,Ando:2004qe,Beacom:2004yd,Berryman:2014qha,Picoreti:2015ika,FRIEMAN1988115,Mirizzi:2007jd,GonzalezGarcia:2008ru,Maltoni:2008jr,Baerwald:2012kc,Broggini:2012df,Dorame:2013lka,Gomes:2014yua,Abrahao:2015rba,Coloma:2017zpg,Choubey:2018cfz,deSalas:2018kri,Aharmim:2018fme,Funcke:2019grs}. 
Mostly model-independent bounds on the decays of $\nu_2$, using the $^8{\rm B}$ solar neutrino data, constrain $\tau_2/m_2 >10^{-3}\,{\rm s/eV}$~\cite{Berryman:2014qha,Aharmim:2018fme}. Similarly, one can use the $^7{\rm Be}$ and low energy $pp$ neutrino data~\cite{Bellini:2014uqa} to constrain $\tau_1/m_1 > 10^{-4}\,{\rm s/eV}$~\cite{Berryman:2014qha}. The tiny magnitude of $\theta_{13}$ makes it very difficult to use solar neutrinos to constrain the $\nu_3$ lifetime. Bounds on $\tau_3$ from atmospheric neutrino data are much weaker, $\tau_3/m_3 >
10^{-10}\,{\rm s/eV}$~\cite{GonzalezGarcia:2008ru,Gomes:2014yua}. Astrophysical bounds arising from the observation of neutrinos from SN1987A~\cite{Kachelriess:2000qc} are quite uncertain, and cannot be treated at the same level of accuracy.

Model-dependent bounds are somewhat stronger. Recently, for example, the authors of \cite{Funcke:2019grs} proposed model-dependent bounds on $\tau_3/m_3 > 2.2\times10^{-5}\,{\rm s/eV}$ using KamLAND solar neutrino data. 
It is also expected that next-generation neutrino oscillation experiments like JUNO~\cite{An:2015jdp} and DUNE~\cite{Acciarri:2016crz} can do a better job constraining invisible $\nu_3$ decays~\cite{Abrahao:2015rba}. Astrophysical neutrinos have longer baselines, and hence can be used to constrain longer lifetimes. 
Constraints from the observation of ultrahigh-energy neutrinos by ICECUBE are potentially stronger, but they suffer from uncertainties in the neutrino flavor composition~\cite{Beacom:2002vi,Bustamante:2016ciw,Denton:2018aml,Abdullahi:2020rge}.
Measurements of the cosmic microwave background can be used to put very strong bounds on the neutrino lifetimes, of order $10^8$~s \cite{Escudero:2019gfk,Chacko:2019nej}, but these are indirect probes of neutrino decay in the sense that one is using information on the expansion rate of the universe to infer whether the neutrino component of the universe in changing.

Typically, neutrino flavor evolution in a supernova is very involved and is marred by the details of collective flavor oscillations. One can circumvent some of these complications by concentrating on the neutrinos emitted in the shock-breakout epoch \cite{Ando:2003ie,Ando:2004qe} that are usually free from such effects. The impact of neutrino decays in current and forthcoming experiments was recently explored by some of the authors in~\cite{deGouvea:2019goq}. 
Depending on the mass-ordering, neutrino decay can either enhance or deplete the neutrino flux in an observable manner.
It was inferred that one can constrain neutrino lifetimes as long as $\tau/m\sim 10^5\,{\rm s/eV} $~\cite{deGouvea:2019goq} from the neutronization-burst neutrino spectra. These bounds are model-dependent, and depend on whether the neutrinos are Dirac or Majorana fermions.

One can extend the same line of argument for the DSNB. Here the typical baseline is even longer than what is expected of an observable individual CCSN. Effects of neutrino decay on the DSNB signal were studied in~\cite{Ando:2003ie,Fogli:2004gy} where it was concluded that such decays could be observable for neutrino lifetimes as long as $\tau/m\sim 10^{10}\,{\rm s/eV}$. The bounds quoted in these works were relatively simple estimates that did not take into account realistic detector-related effects, including systematics and the effect of finite detector energy resolution. In this section, we revisit the sensitivity of measurements of the DSNB to the neutrino lifetime and calculate the signals that can be observed in current and near-future generation detectors. The specifics are, of course, model-dependent, and also depend on whether neutrinos are Dirac or Majorana fermions~\cite{deGouvea:2019goq}. 
For concreteness, we consider a scenario where the neutrinos are Majorana fermions and interact with an almost-massless scalar $\varphi$. The interaction is described by
\begin{equation}\label{eq:DecLag}
\mathcal{L}\supset \frac{f_{ij}}{2} (\nu_L)_i (\nu_L)_j \varphi + {\rm H.c.}\,.
\end{equation}

At the tree level, this operator allows a heavy neutrino $\nu_j$ to decay into a lighter neutrino $\nu_i$, and $\varphi$. For concreteness, we consider the scenario where the heaviest neutrino decays into the lightest one, i.e., if the neutrino mass ordering is normal (inverted), a $\nu_3\,(\nu_2)$ decays into a $\nu_1\,(\nu_3)$. Although the neutrino mass ordering is still not known, global fits of neutrino oscillation data favor the normal mass ordering \cite{Esteban:2018azc} (see, however,~\cite{Kelly:2020fkv}), which will be assumed throughout. Results for the inverted mass ordering are qualitatively similar. 

Assuming the $\nu_1$ is massless, the rate of decay is given by 
\begin{equation} \label{eq:rate}
 \Gamma(E_3)= 2\times \frac{f^2 m_3^2}{64\pi E_3}\,,
\end{equation}
where the factor of 2 arises because both the left helicity (L) and right helicity (R) final states contribute. Depending on whether the daughter neutrinos have the same helicity as the parent (helicity-conserving--h.c.) or different helicity (helicity-flipping--h.f.), the energy distribution of the daughter particles is,
\begin{eqnarray} \label{eq:energydist}
 \psi_{\rm h.c.}(E_3,E_1)\equiv\frac{1}{\Gamma}\,\frac{d\,\Gamma}{d\,E_1}& \propto&\frac{2 E_1}{E_3^2}\qquad\qquad\qquad\,\,{\rm for\,\,} \nu_{3_L} \to \nu_{1_L} + \varphi_0\,\,, \nonumber \\
 \psi_{\rm h.f.}(E_3,E_1)\equiv\frac{1}{\Gamma}\,\frac{d\,\Gamma}{d\,E_1}& \propto&\frac{2}{E_3}\left(1-\frac{E_1}{E_3}\right)\qquad{\rm for\,\,} \nu_{3_L} \to \nu_{1_R} + \varphi_0\,\,.
 \end{eqnarray}
Since we are assuming the neutrinos are Majorana neutrinos, the ``wrong" helicity neutrinos act as what we typically refer to as ``antineutrinos'' (and vice-versa), and will interact with the detector even in the limit where the daughter neutrino is massless. On the other hand, for Dirac neutrinos, the ``wrong"-helicity neutrinos (right-handed neutrino states and left-handed antineutrino states) are essentially inert.
 
To calculate the daughter $\nu_1$ flux from the decay of the $\nu_3$, one needs to compute a Boltzmann transfer equation for $\nu_1$, involving terms which can create or destroy $\nu_1$. The flux of the mass eigenstates arriving at the Earth is 
\begin{eqnarray}
\Phi_{\nu_3}(E)&=&\int_0^{z_{\rm max}}\ \frac{dz'}{H(z')}\,\R(z')\,F_{\nu_3}\left(E(1+z')\right)e^{-\Gamma(E)\zeta(z') }\nonumber\\
\Phi_{\nu_2}(E)&=&\int_0^{z_{\rm max}}\ \frac{dz'}{H(z')}\,\R(z')\,F_{\nu_2}\left(E(1+z')\right) \nonumber\\
\Phi_{\nu_1}(E)&=&\int_0^{z_{\rm max}}\ \frac{dz'}{H(z')}\R(z')\,F_{\nu_1}\left(E(1+z')\right)+\nonumber\\     
            &&\int_0^{z_{\rm max}}\ \frac{dz'}{H(z')} \int_{E(1+z')}^\infty\, dE'\, \bigl[\Phi_{\nu_3}(E')\,\Gamma(E') \, \psi_{\rm h.c.}(E',E(1+z')) + \Phi_{\bar{\nu}_3}(E')\,\Gamma(E') \, \psi_{\rm h.f.}(E',E(1+z'))\bigr]\,,\notag\\
\end{eqnarray}
where $\zeta(z)=\int_0^z dz'\, H^{-1}(z')(1+z')^{-2}$, and $z_{\rm max}=5$. Similar expressions hold for the antineutrinos. For a detailed derivation of these transfer equations, see \cite{Fogli:2004gy}.

\begin{figure}[!t]
\includegraphics[width=.75\textwidth]{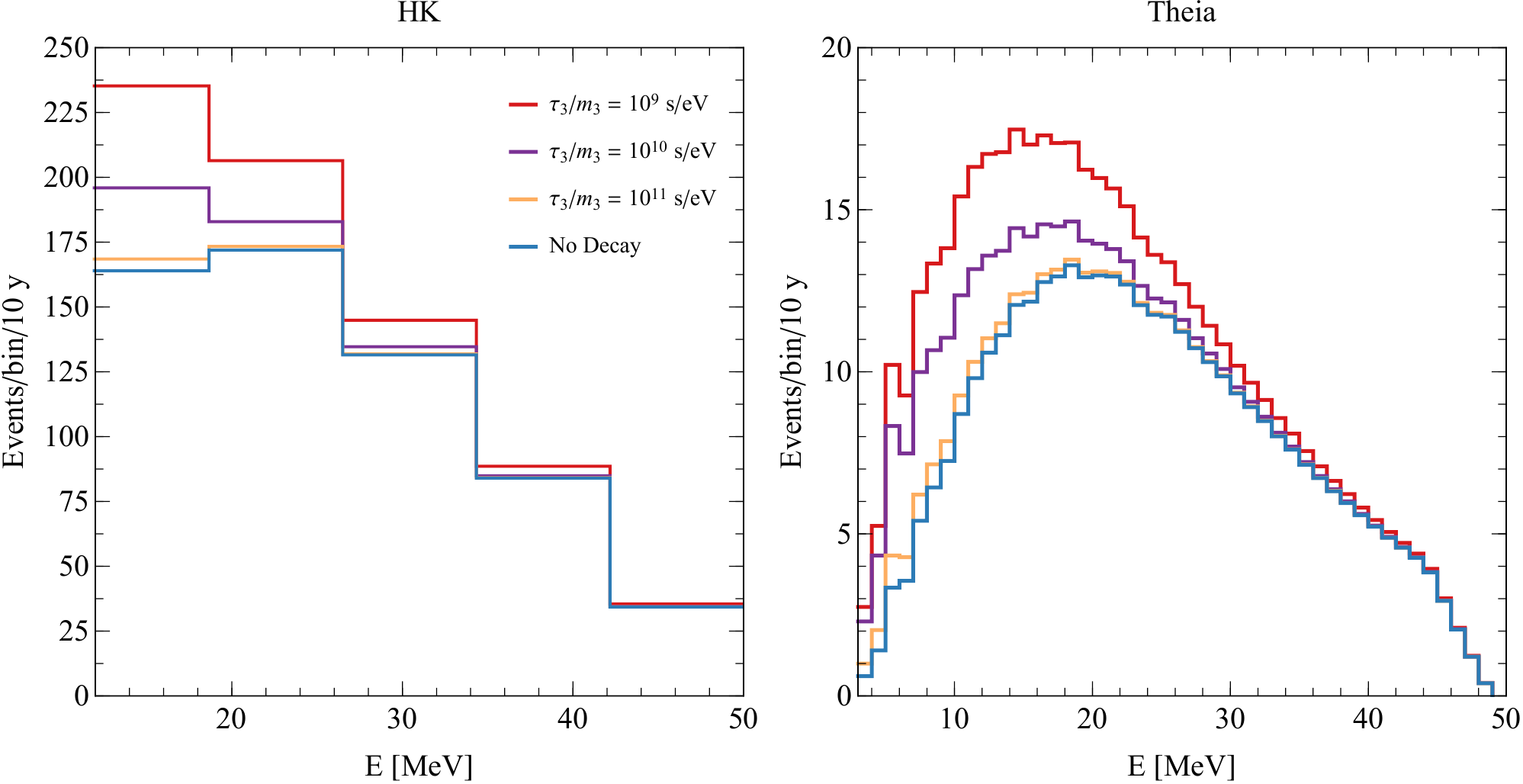}
\caption{ Expected event rates as a function of the reconstructed positron energy after ten years of data-taking in HK (left) and Theia (right) for different values of the heaviest neutrino lifetime. See text for details.}
\label{fig:EventDecay}
\end{figure}

Using this formalism, we explore the consequences of neutrino decay on the DSNB flux. 
Fig.\,\ref{fig:EventDecay} depicts the expected neutrino spectra in HK (left) and Theia (right) for three different lifetimes for the heaviest mass eigenstate, $\nu_3$. As expected, the number of events in each bin increases if the neutrino decays faster. While HK profits from larger statistics, Theia has better energy resolution and is more sensitive to distortions in the spectral shape.

\begin{figure}[!t]
\includegraphics[width=.4\textwidth]{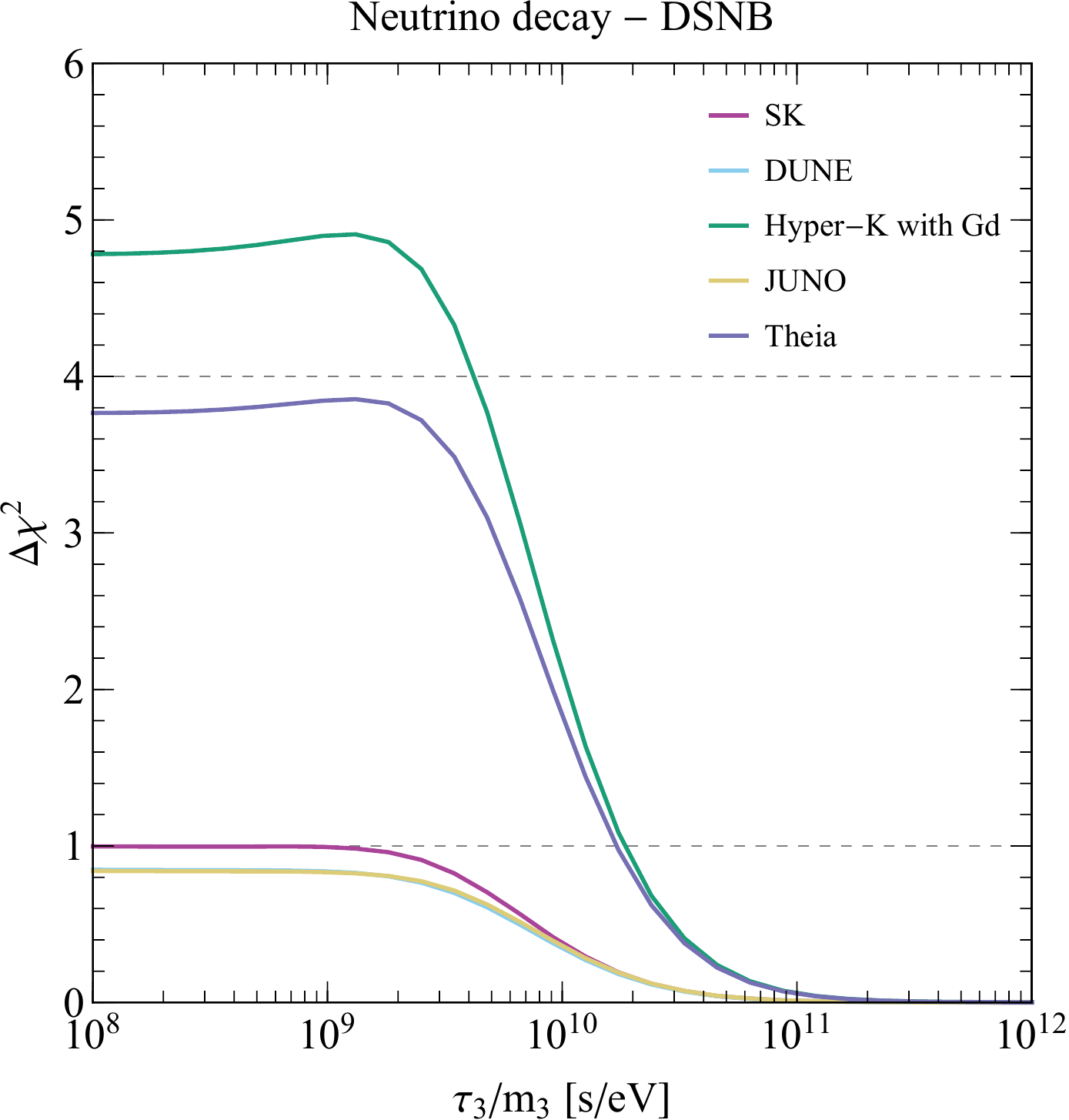}
\caption{$\Delta\chi^2$ (relative to the minimum value) as a function of the $\nu_3$ lifetime-over-mass $\tau_3/m_3$, for ten years of simulated data at different current and future experiments.}
\label{fig:DecayDSNB}
\end{figure}

We consider the data to be consistent with the hypothesis that the neutrinos do not decay, and perform a simple $\chi^2$ analysis to estimate the bounds forthcoming experiments can set on the neutrino lifetime. Fig.\,\ref{fig:DecayDSNB} depicts $\Delta \chi^2$ as a function of the lifetime over mass $\tau_3/m_3$ for some of the current and forthcoming experiments sensitive to the DSNB. Near-future experiments like SK and JUNO (and DUNE), are expected to observe only a few events and hence are not very sensitive to a finite $\nu_3$ lifetime. HK (doped with Gd) and Theia, on the other hand, are sensitive, at more than the two-sigma level, to lifetimes as long as $10^{9}\,{\rm s/eV}$. In the fit, we assume the parameters associated with the SFR, the time evolution of the universe, and neutrino mixing are known perfectly and agree with their current best fit value. As discussed earlier, the 40\% systematic uncertainty assumed for the overall neutrino flux partially takes care of uncertainties associated to the SFR, cosmology, and neutrino mixing. 
This sensitivity, while somewhat limited, rivals some of the best bounds on neutrino lifetimes, and would complement bounds inferred indirectly from cosmology~\cite{Escudero:2019gfk,Chacko:2019nej}.  
The lack of higher statistics prevents us from getting a higher sensitivity to neutrino lifetimes even in these experiments. Note that these sensitivity estimates are around one order of magnitude smaller than those quoted in \cite{Fogli:2004gy} that are less detailed when it comes to more specific experimental issues, including backgrounds and the energy resolution at different future experiments. More statistics leads to better sensitivity. After 20 years of data taking, for example, we estimate that HK can reach three-sigma sensitivity to $\tau_3/m_3\lesssim 5\times 10^{9}$~s/eV. 

\subsection{Probing pseudo-Dirac neutrinos with the DSNB}
\label{sec:pseudo}
Neutrinos are the only Standard Model fermions whose fundamental nature is still unknown. They can be either Dirac or Majorana particles, depending on whether total lepton number is exactly conserved in nature. An intriguing possibility is that neutrinos are Majorana fermions but the lepton-number breaking structure of the theory is such that neutrinos behave, for the most part, as Dirac fermions~\cite{Wolfenstein:1981kw,Petcov:1982ya,Bilenky:1983wt,Kobayashi:2000md,deGouvea:2009fp}. Under these circumstances, neutrinos are referred to as pseudo-Dirac fermions. 

If neutrinos are pseudo-Dirac fermions, right-handed neutrino fields $N_i$ ($i=1,2,\ldots$, at least two of them) exist and couple of the Standard Model lepton doublets $L_{\alpha}$ ($\alpha=e,\mu,\tau$) and the Higgs double $\Phi$ via Yukawa interactions (proportional to Yukawa couplings $Y_{i\alpha}$). In the absence of more sources of mass, after electroweak symmetry breaking, neutrinos are Dirac fermions with mass $M_D=Yv/\sqrt{2}$, where $v/\sqrt{2}$ is the vacuum expectation value of the neutral component of the Higgs doublet. If, along with $M_D$, there is another source for the neutrino masses -- in the form of either Majorana masses $M_R$ for the right-handed neutrinos or Majorana masses for the left-handed neutrinos -- these would arise, after integrating out new, heavy degrees of freedom, via the Weinberg operator, $(L\Phi)^2$ -- and the new source of mass is subdominant to $M_D$, neutrinos are pseudo-Dirac fermions.  

For concreteness, we concentrate on the so-called Type-I seesaw scenario, where, after electroweak symmetry breaking, the neutrino Majorana mass Matrix $M_{ij}$ ($i,j=1,2,3,\ldots, 3+n_R$, where $n_R$ is the number of right-handed neutrino fields), is 
\begin{align}\label{eq:PDLag}
    M=\begin{pmatrix}
        \mathbb{0}_3 & M_D\\
        M_D & M_R
    \end{pmatrix}\,,
\end{align}
where $\mathbb{0}_3$ is a $3\times 3$ matrix of zeros. In the pseudo-Dirac regime, $M_R\ll M_D$ and we will assume $n_R$ is equal to three henceforth. 

It is straight forward to diagonalize the $6\times 6$ mass matrix in Eq.\ \eqref{eq:PDLag} via $V^\dagger (M^\dagger M) V = M_{\rm diag}^2$ perturbatively in this regime. The unitary $6\times 6$ matrix $V$ can be written as \cite{Kobayashi:2000md}
\begin{align}
    V=\frac{1}{\sqrt{2}}\begin{pmatrix}
    U & 0\\
    0 & U_R
    \end{pmatrix}\cdot
    \begin{pmatrix}
    \mathbb{1}_3 & i \mathbb{1}_3\\
    \varphi & -i\varphi
    \end{pmatrix}\,,
\end{align}
where $\mathbb{1}_3$ is the $3\times 3$ identity matrix, $U$ is a $3\times 3$ unitary matrix that is equivalent of the leptonic mixing matrix in the limit $M_R\to 0$, $U_{R}$ is a (unphysical) $3\times 3$ matrix which diagonalizes the ``sterile'' sector mass matrix, and $\varphi$ depends on unknown phases $\varphi = {\rm diag}(e^{-i\phi_1},e^{-i\phi_2},e^{-i\phi_3})$. The active neutrinos can be expressed as linear superpositions of the six mass eigenstates, $\{\nu_{js},\nu_{ja}\}, j=1,2,3$:
\begin{align}
    \nu_{\alpha L}&=\frac{1}{\sqrt{2}}U_{\alpha j}(\nu_{js}+i\,\nu_{ja})\,.
\end{align}
These, in turn, have masses-squared
\begin{align}
    m_{ks}^2 = m_k^2 + \frac{1}{2}\delta m_k^2,\quad  m_{ka}^2 = m_k^2 - \frac{1}{2}\delta m_k^2\,,
\end{align}
$k=1,2,3$. $m_k$ and $\delta m_k^2$ are functions of the elements of $M_D$ and $M_R$. In the limit $M_R\to 0$, $\delta m_k^2\to 0$ and $m_k^2$ are the eigenvalues of $M_DM_D^\dagger$. In this limit, it is easy to see that the neutrinos are Dirac fermions. 

The pseudo-Dirac limit, $M_R\ll M_D$, can be understood as follows. Neutrino mass eigenstates are organized into three pairs of quasi-degenerate states, separated by small mass-squared differences $\delta m^2_k$, which are much smaller than the known ``solar'' $\Delta m^2_{21}$ and ``atmospheric'' $|\Delta m^2_{31}|$. These degenerate pairs are fifty-fifty linear combinations -- maximal mixing -- of active and sterile neutrinos. These small mass-squared differences will lead to oscillations at very large distances. Limits on this scenario are obtained by considering solar neutrino oscillations, sensitive to mass-squared differences down to $\delta m_k^2\sim10^{-12}\eV^2$~\cite{deGouvea:2009fp,Donini:2011jh}, and from atmospheric neutrinos, affected if $\delta m_k^2\gtrsim 10^{-4}\eV^2$~\cite{Beacom:2003eu}. High-energy astrophysical neutrinos constrain $10^{-18}\eV^2\lesssim \delta m_k^2\lesssim 10^{-12}\eV^2$, but these bounds depend strongly on knowledge of the Gamma Ray Burst mechanism and the neutrino energy~\cite{Beacom:2003eu,Esmaili:2012ac}. 

The DSNB was emitted, on average, at large redshifts and hence allows one to probe baselines of order a few~Gpc. Therefore, we expect sensitivity to very tiny, new mass-squared differences. Oscillation probabilities can be obtained in the standard way~\cite{Kobayashi:2000md,deGouvea:2009fp}, with a few extra ingredients that are worthy of note. The initial states are, according to our oscillation discussion in Section~\ref{sec:model}, the ``active'' mass eigenstate $\nu_{ia}$, $i=1,2,3$, which will be detected as a $\nu_{\alpha}$ on Earth with probability $|U_{\alpha i}|^2$ (ignoring Earth matter-effects). These are the states that oscillate into sterile states (with maximal mixing). The oscillation phase has to be corrected in order to include the energy redshift since neutrinos are propagating in an expanding Universe~\cite{Esmaili:2012ac,Beacom:2003eu}. Furthermore, given the cosmological distances, wave-packet-separation decoherence effects may also play an interesting role and are included.

The oscillation probability for $\nu_i\to\nu_\beta$, as function of the neutrino energy today and the redshift, is
\begin{align}
    P_{i\beta}(z,E) = \frac{1}{2}|U_{\beta k}|^2\left(1+e^{-\mathscr{L}_k^2(z)}\cos\left(\frac{\delta m_k^2}{2E}L_2(z)\right)\right)
\end{align}
where the decoherence factor $\mathscr{L}_k(z)$ is~\cite{Giunti:1997wq}
\begin{align}
    \mathscr{L}_k(z) &=\frac{\delta m_k^2}{4\sqrt{2}E^2}\frac{L_3(z)}{\sigma_x}
\end{align}
and $\sigma_x$ the initial size of the wave packet. The distances $L_n(z)$ are ~\cite{Esmaili:2012ac,Beacom:2003eu}
\begin{align}
     L_n(z)&= \frac{1}{H_0}\int_0^z\frac{dz^\prime}{(1+z^\prime)^n}\frac{1}{\sqrt{\Omega_m(1+z^\prime)^3+\Omega_\Lambda}}\notag\\
     &=\frac{1}{H_0}\left\{ _2F_1\left[1,(5-2n)/6;(4-n)/3;-\Omega_m/\Omega_\Lambda\right]-(1+z)^{1-n}\, _2F_1\left[1,(5-2n)/6;(4-n)/3;-(1+z)^3\Omega_m/\Omega_\Lambda\right]\right\},
\end{align}
where $ _2F_1[a,b;c;z]$ is the hypergeometric function. 

Oscillation and decoherence effects are important for mass-squared differences and wave-packet sizes of order
\begin{subequations}
\begin{align}
    L_{\rm osc} &= \frac{4\pi E}{\delta m_k^2} \approx 8.03\Gpc\left(\frac{E}{10\MeV}\right)\left(\frac{10^{-25}\eV^2}{\delta m_k^2}\right),\\
    L_{\rm coh} &= \frac{4\sqrt{2} E^2}{|\delta m_k^2|}\sigma_x \approx 180\Gpc \left(\frac{E}{10\MeV}\right)^2\left(\frac{10^{-25}\eV^2}{\delta m_k^2}\right)\left(\frac{\sigma_x}{10^{-12}{\rm\ m}}\right).
\end{align}
\end{subequations}
Hence, measurements of the DSNB are sensitive to mass-squared differences as small as $\mathscr{O}(10^{-25}\eV^2)$. For mass-squared differences $\delta m_k^2\lesssim 10^{-28}\eV^2$, the naive oscillation length is larger than the size of the observable Universe and, for neutrino energies $E\sim \mathscr{O}(10\MeV)$, such mass-squared differences are simply not accessible. 

Decoherence effects may be significant if $\sigma_x$ is small enough. For an initial wave packet size of $\sigma_x\sim 1{\rm\ fm}$ and $\delta m^2_k\sim 10^{-25}~{\rm \eV}^2$, for example, the DSNB neutrinos arrive at the Earth as incoherent superpositions of active and sterile states. In this case, the expect flux is 50\% that of not-pseudo-Dirac neutrinos. In fact, if the $\sigma_x$ values were of order one fermi, decoherence effects would always supersede oscillation effects for typical DSNB energies,
\begin{align*}
    \frac{L_{\rm coh}}{L_{\rm osc}}=\frac{\sqrt{2}}{\pi}E\sigma_x\approx 0.0023\left(\frac{E}{10\MeV}\right)\left(\frac{\sigma_x}{10^{-15}{\rm\ m}}\right)\,.
\end{align*} 
In this regime, it is unfair to refer to the phenomenon in question as quantum mechanical oscillations. Incoherent mixtures of active and sterile states would be a more appropriate description. For the rest of this discussion, we will be agnostic concerning the value of $\sigma_x$. We assume that it cannot be smaller than one fermi -- the size of a proton -- given how CCSN neutrinos are produced and the conditions under which they are produced. In fact, we expect  $\sigma_x\gg 1$~fm; thoughtful studies in the literature have argued for $\sigma_x\sim 10^{-13}$~m~\cite{Kersten:2015kio}.

\begin{figure}[!t]
\includegraphics[width=0.75\textwidth]{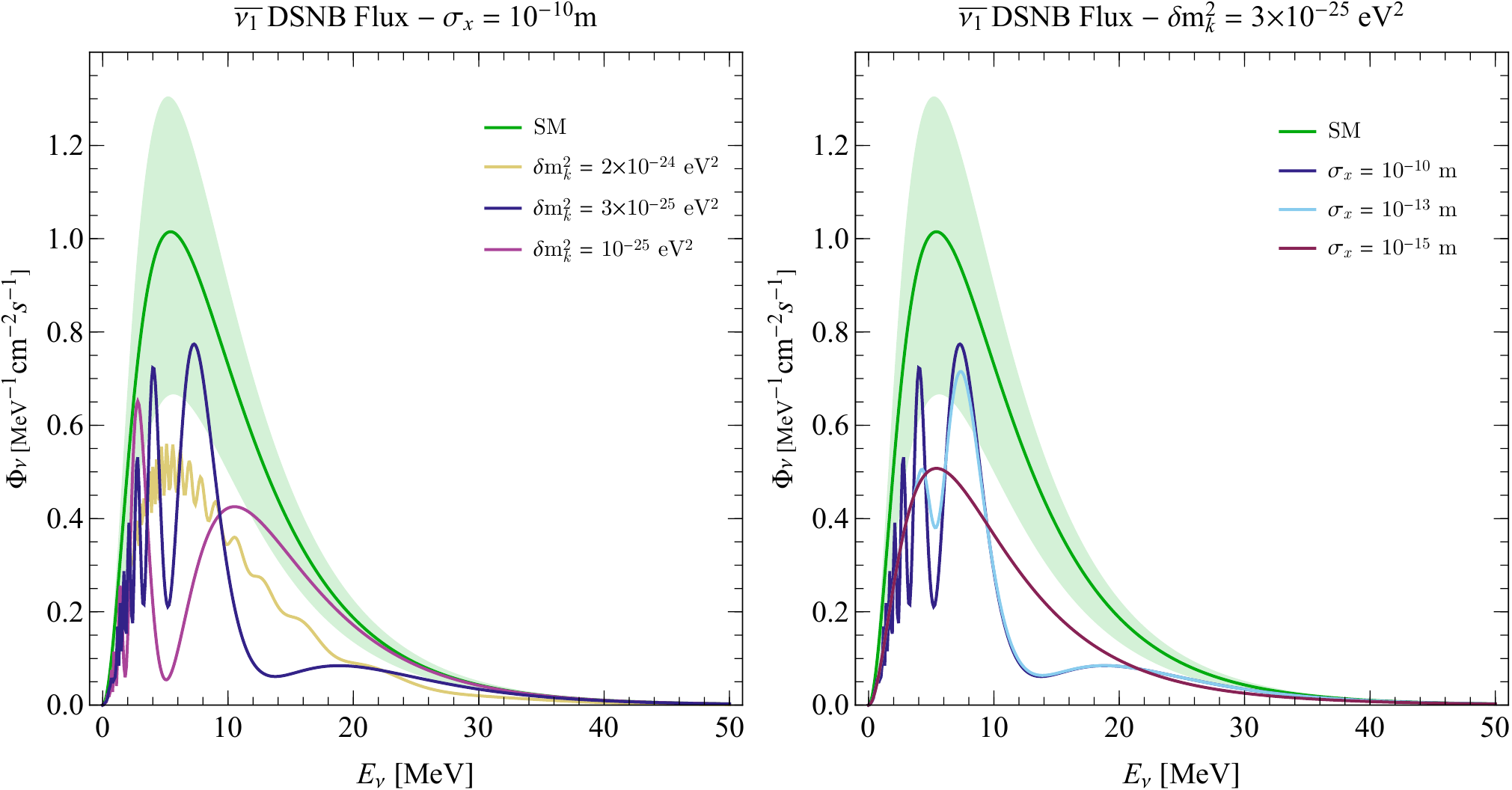}
\caption{The impact of active--sterile oscillations on the expected active DSNB $\bar{\nu}_1$ flux, assuming the neutrinos are pseudo-Dirac fermions. The left panel includes different values of $\delta m_k^2$ for a fixed value of $\sigma_x=10^{-10}{\rm\ m}$. The right panel includes different values of the wave-packet size $\sigma_x$ for a fixed value of $\delta m_k^2=3\times 10^{-25}\eV^2$.The green shaded region indicates the modeling uncertainty on the normalization of the DSNB.}
\label{fig:osciflux}
\end{figure}

Fig.~\ref{fig:osciflux} depicts the modification of the DSNB active $\bar{\nu}_1$ flux due to active-sterile oscillations in the pseudo-Dirac scenario. In the left panel, $\sigma_x=10^{-10}$~m is held fixed and the different curves correspond to different values of $\delta m_k^2$. We observe an oscillatory behavior for the active flux on Earth for $\delta m^2_k\lesssim 10^{-24}$~eV$^2$. For larger values, the oscillations average out and one is left with half of the unoscillated flux. In the right panel, $\delta m_k^2=3\times 10^{-25}\eV^2$ is held fixed and the different curves correspond to different values of $\sigma_x$. 

\begin{figure}[!t]
\includegraphics[width=\textwidth]{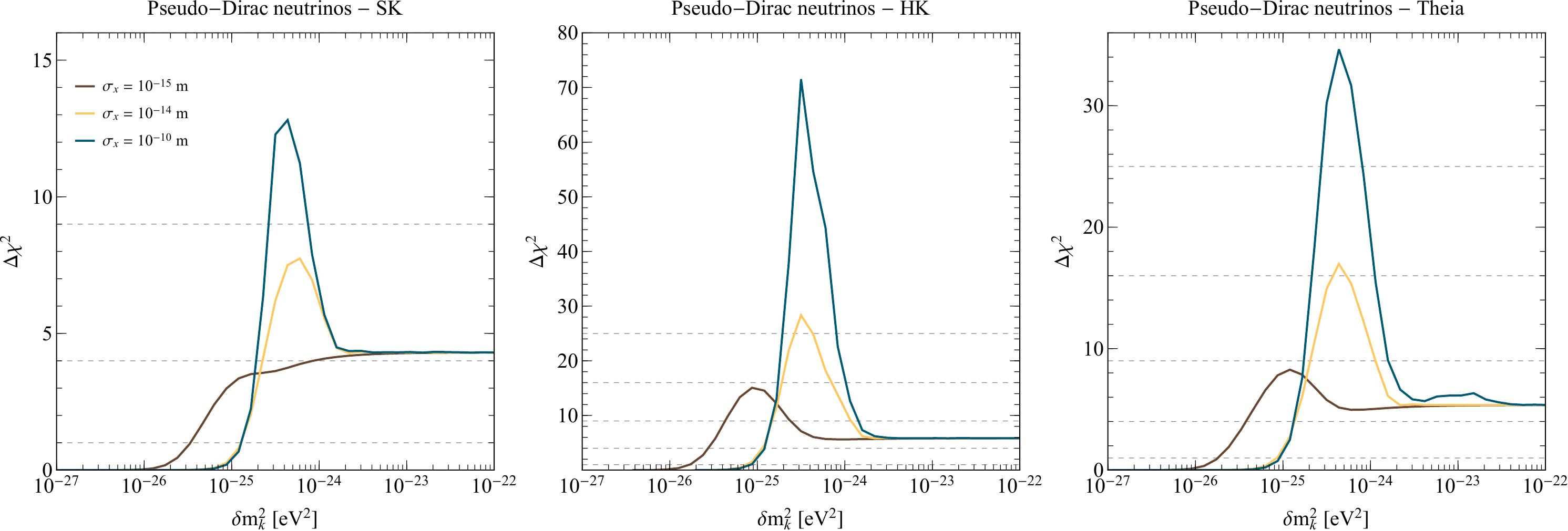}
\caption{$\Delta\chi^2$ (relative to the minimum value) as a function of $\delta m_k^2$, for ten years of simulated data at SK (left), HK (middle), and Theia (right), for three different values of the decoherence parameter $\sigma_{x} = (10^{-15}\,{\rm m}, 10^{-14}\,{\rm m},10^{-12}\,{\rm m})$.}
\label{fig:pd1d}
\end{figure}

Assuming data consistent with no active-sterile oscillations, we perform a $\chi^2$ analysis asking how future facilities could constrain the new mass-squared differences $\delta m_k^2$, for fixed values of the initial size of the wave-packet. Here, we assume the three new mass-squared difference are the same, $\delta m^2_{1}=\delta m^2_{2}=\delta m^2_{3}$. Fig.\ \ref{fig:pd1d} depicts the sensitivity to the pseudo-Dirac scenario for SK, (left), HK (center) and Theia (right). When the decoherence effects are not important, i.e., for $\sigma_x=10^{-10}{\rm\ m}$, SK is sensitive to a new mass-squared difference at the $2\sigma$ C.L.\ if  $\delta m_k^2\gtrsim 1.8\times 10^{-25}\eV^2$, while HK (Theia) would be sensitive to $\delta m_k^2\gtrsim 10^{-25}\eV^2$ ($\delta m_k^2\gtrsim 1.3\times 10^{-25}\eV^2$). Moreover, since the new oscillations would induce a large modification of the shape of the fluxes in the observable region, we find a significant enhancement in the sensitivity for  $10^{-25}\eV^2 \lesssim \delta m_k^2\lesssim 10^{-24}\eV^2$, reaching values of $\Delta\chi^2=70~(30)$ for HK (Theia). Ten years of SK data can rule out $\delta m_k^2\in \{2.5,7.9\} \times 10^{-24}\eV^2$ at the three sigma level. For values of $\delta m_k^2\lesssim 10^{-26}\eV^2$, the oscillation lengths are larger than the cosmological baselines of the DSNB, so the sensitivity is effectively lost. On the other hand, if the mass-squared difference is large, $\delta m_k^2 \gtrsim 10^{-24}\eV^2$, the oscillations are averaged out, and neutrinos arrive basically as incoherent superpositions of active and sterile states. The sensitivity in that case does not depend on $\delta m^2_k$, plateauing at $\Delta\chi^2 = \{4.3,5.9,5.5\}$ for SK, HK and Theia, respectively. 
\begin{figure}[!t]
\includegraphics[width=\textwidth]{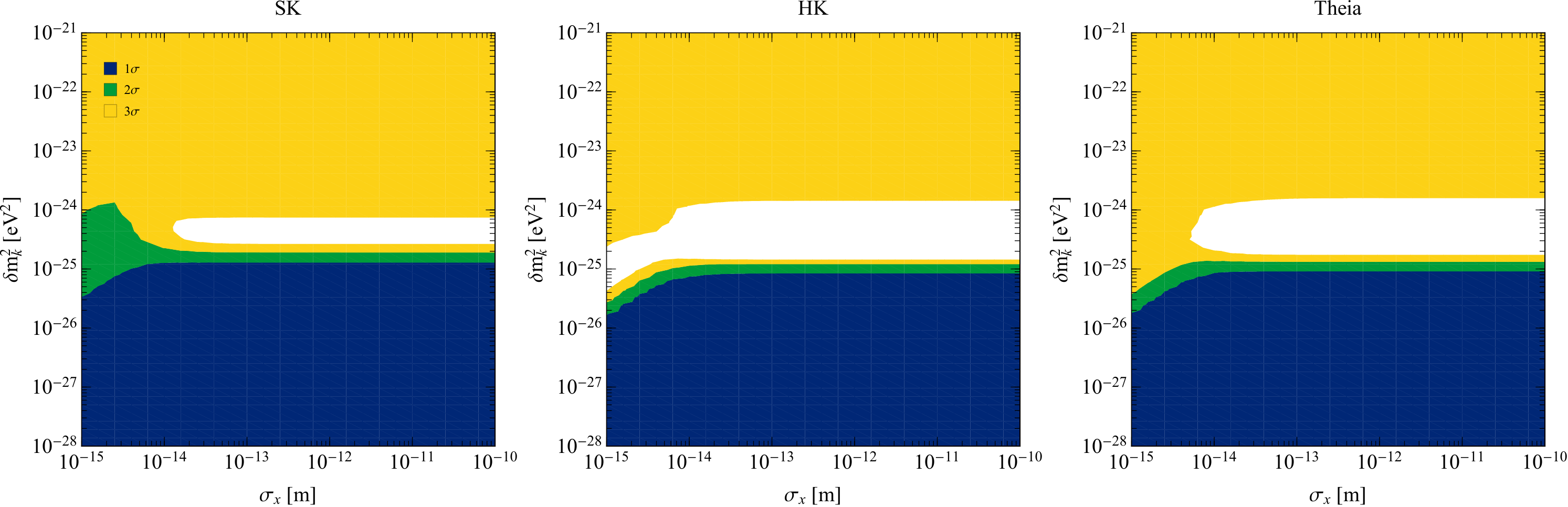}
\caption{Sensitivity on the  ($\sigma_{x}$, $\delta m^2_k$) parameter space from ten years of simulated data at SK (left), HK (middle), and Theia (right). The regions that can be tested at less than or equal to $1\sigma, 2\sigma, 3\sigma$ level correspond to the blue, green and yellow areas, respectively. The white regions have a sensitivity $>3\sigma$ C.L.}
\label{fig:pd2d}
\end{figure}

One interesting effect appears when the decoherence due to the initial wave packet size is important. Since the oscillations are erased by the separation of the wave packets, the modification of the DSNB spectra becomes less pronounced, thus reducing the overall sensitivity for all three experiments considered. However, we find that the sensitivity is shifted to lower values of $\delta m_k^2$ because the decoherence effects are present even for cases in which the oscillations would be absent. The effect is maximal for $\delta m_k^2\sim 10^{-25}\eV^2$ and $\sigma_x=10^{-15}{\rm\ m}$ in all experiments considered since, for these values of the relevant parameters, decoherence effects impact the shape of the DSNB. This enlarges the accessible region of parameter space, and one is sensitive to  $\delta m_k^2\sim 5\times 10^{-25}\eV^2$ for HK and Theia at the $2\sigma$ C.L., while for SK the sensitivity remains below the $2\sigma$ level for $\delta m_k^2\lesssim 10^{-24}\eV^2$. The sensitivity to pseudo-Dirac neutrinos as a function of $\sigma_x$ and $\delta m^2_k$ is depicted in Fig.\ \ref{fig:pd2d}  for SK, (left), HK (center) and Theia (right). We observe that the sensitivity is independent of $\sigma_x$ for $\sigma_x\gtrsim 10^{-14}{\rm\ m}$ at the three experiments. For some range of mass-squared differences, all experiments can rule out the pseudo-Dirac neutrino hypothesis at more than the three-sigma level as long as $\sigma_x$ is large enough. In the case of HK, even when decoherence effects are strong, one can rule out  the pseudo-Dirac neutrino hypothesis at more than the three-sigma level for some range of mass-squared differences.

\section{Conclusions}
\label{sec:conc}
%
%


Since the dawn of the first stars, core-collapse supernovae have created a plethora of tens-of-MeV neutrinos of all species.
This steady source of CCSN relic neutrinos, the diffuse supernova neutrino background (DSNB), serves as a complementary probe of certain aspects of cosmology, astrophysics, and particle physics. As Super-Kamiokande, doped with gadolinium, prepares to take data, the detection of the DSNB appears to be just a matter of time. As a result, a detailed study of the role played by the DSNB in this era of multi-messenger astronomy is timely. This is what we pursued here. 

Predicting the DSNB flux requires inputs from three different disciplines: (i) the cosmological history of the Universe, (ii) the rate of CCSNe, which depends on the star-formation rate and is constrained with astrophysical data and modeling, and (iii) the neutrino energy spectra from  CCSNe that depend on simulations of CCSNe and the particle-physics properties of neutrinos. As a result, successful detection of the DSNB in currently running and future detectors can help to constrain different cosmological parameters, the star-formation history, as well as different types of beyond-the-Standard-Model particle physics. In this work, we have analyzed the potential of terrestrial detectors to constrain the above, using the DSNB.

\begin{figure}[!t]
\includegraphics[width=0.45\textwidth]{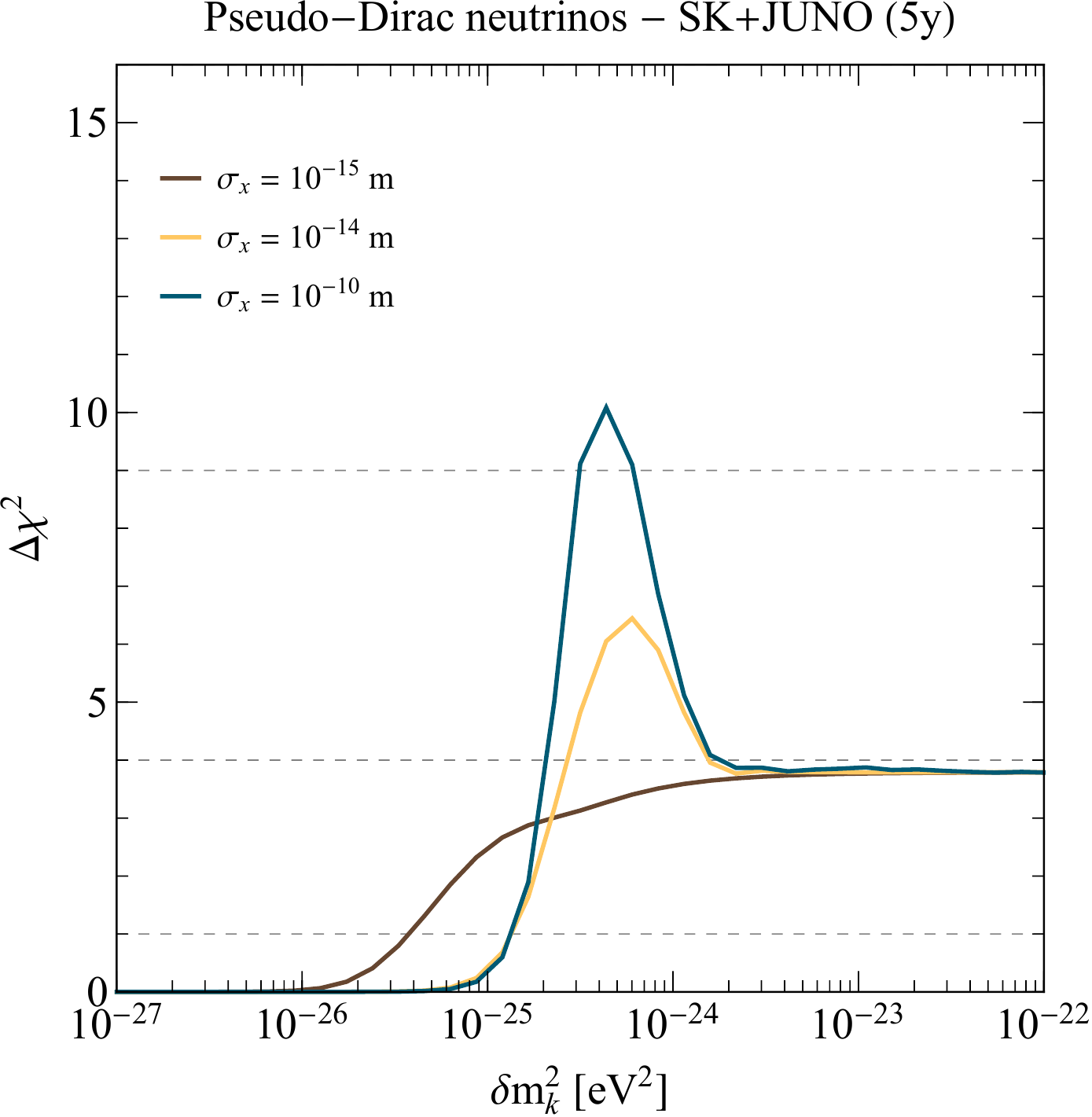}\qquad\qquad
\includegraphics[width=0.405\textwidth]{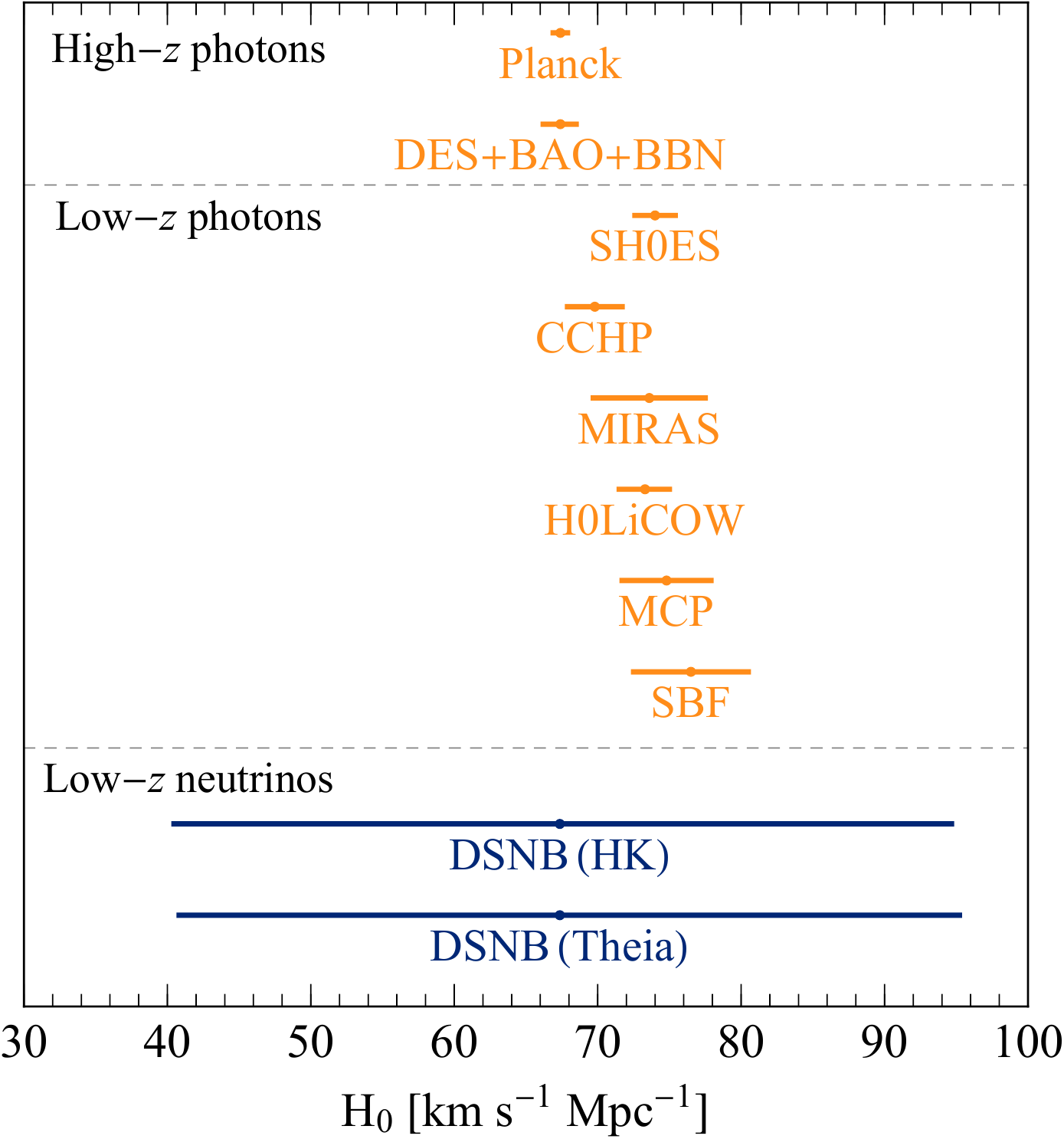}
\caption{Left: $\Delta\chi^2$ (relative to the minimum value) as a function of $\delta m_k^2$, for five of years of combined simulated data at SK and JUNO, for three different values of the decoherence parameter $\sigma_{x} = (10^{-15}\,{\rm m}, 10^{-14}\,{\rm m},10^{-10}\,{\rm m})$. Right: current measurements of the Hubble constant $H_0$ from Planck \cite{Aghanim:2018eyx},
DES+BAO+BBN \cite{2018MNRAS.480.3879A},
SH0ES \cite{Riess:2016jrr,Riess:2019cxk},
CCHP \cite{Freedman:2019jwv},
MIRAS \cite{Huang:2019yhh},
H0LICOW \cite{Wong:2019kwg},
MCP \cite{2009ApJ...695..287R}, and
SBF \cite{2018AAS...23231902P},
along with our estimates for the precision with which $H_0$ can be extracted from future DSNB data.} 
\label{fig:pd1d_now}
\end{figure}

In ten years, JUNO and SK are expected to accumulate a few to several tens of events. In ten years, HK and Theia should have hundreds of events in hand. These samples dictate what information one can hope to extract from measurements of the DSNB. Before HK and Theia are available, combined SK and JUNO can already yield interesting results. In Fig.~\ref{fig:pd1d_now}(left) we depict how well combined JUNO and SK data, collected over a five-year period, can constrain the hypothesis that neutrinos are pseudo-Dirac fermions. We find that a small range of new mass-squared differences (around several times $10^{-25}$~eV$^2$) can be ruled out at the three-sigma level. We refer to Sec.~\ref{sec:pseudo} for more details. 

More statistics allow one to consider more ambitious measurements. We argued that HK and Theia, after a decade of running, can provide a ``neutrino-measurement'' of the expansion rate of the universe. Fig.~\ref{fig:pd1d_now}(right) depicts the current ``electromagnetic'' measurements of the Hubble constant at both low and high redshift, along with our estimate for how precisely HK and Theia can measure $H_0$ after ten years of DSNB data are collected. The latter are expected to be systematics dominated; improved estimates of the star-formation-rate allow for more precise measurements of $H_0$. While it is clear that DSNB measurements of $H_0$ will not compete, precision-wise, with current measurements of $H_0$ -- they probably will not be able to seriously inform the current tension between low- \cite{Riess:2016jrr,Riess:2019cxk,Freedman:2019jwv,Huang:2019yhh,Wong:2019kwg,2009ApJ...695..287R} and high-redshift  data \cite{Aghanim:2018eyx,2018MNRAS.480.3879A,2018AAS...23231902P} -- it is also clear that a neutrino-measurement of the expansion of the Universe would be invaluable. Given expectations that $H_0$ will be measured using the detection of gravity-waves from the collisions of ultra-dense objects \cite{Schutz:1986gp,Holz:2005df,Chen:2017rfc}, it seems we are only a decade away from doing multi-messenger cosmology with photons, gravitons, and neutrinos!

\section*{Acknowledgements}
We thank Boris Kayser for illuminating discussions on decoherence. 
The work of AdG is supported in part by the DOE Office of Science
award \#DE-SC0010143.
MS acknowledges support from the National Science Foundation, Grant PHY-1630782, and
to the Heising-Simons Foundation, Grant 2017-228. Fermilab is operated by the Fermi Research Alliance, LLC under contract No. DE-AC02-07CH11359 with the United States Department of Energy.

\bibliographystyle{kpmod}
\bibliography{DSNB}
\end{document}